\documentclass[11pt]{article}
\usepackage{graphicx,ifthen,color,algorithm,palatino}
\usepackage{amsmath,amsthm,amssymb,amsfonts,verbatim}
\usepackage{mathrsfs,accents,setspace}
\usepackage{subcaption}
\usepackage[autostyle]{csquotes}
\newboolean{DoubleSpaced}
\setboolean{DoubleSpaced}{false}
\definecolor{ruppertgreen}{rgb}{0,.7,.25}

\def\bib{\vskip12pt\par\noindent\hangindent=1 true cm\hangafter=1}
\def\ynew{y_{\mbox{{\tiny new}}}}
\def\tr{\mbox{tr}}
\def\myand{\&\ }
\def\balpha{\boldsymbol{\alpha}}
\def\real{{\mathbb R}}
\def\diag{\mbox{diag}}
\def\simind{\stackrel{{\tiny \mbox{ind.}}}{\sim}}
\def\bzero{\boldsymbol{0}}
\def\Ksc{{\mathcal K}}
\def\smhalf{{\textstyle{\frac{1}{2}}}}
\def\bsigma{\boldsymbol{\sigma}}
\def\ba{\boldsymbol{a}}
\def\rest{\mbox{\rm rest}}
\def\const{\mbox{\rm const}}
\def\bC{\boldsymbol{C}}
\def\bmu{\boldsymbol{\mu}}
\def\bone{\boldsymbol{1}}
\def\btheta{\boldsymbol{\theta}}
\def\bc{\boldsymbol{c}}
\def\bbeta{\boldsymbol{\beta}}
\def\bu{\boldsymbol{u}}
\def\bv{\boldsymbol{v}}
\def\bM{\boldsymbol{M}}
\def\bI{\boldsymbol{I}}
\def\bSigma{\boldsymbol{\Sigma}}
\def\bX{\boldsymbol{X}}
\def\by{\boldsymbol{y}}
\def\bZ{\boldsymbol{Z}}
\def\sumin{\sum_{i=1}^n}
\def\Var{\mbox{Var}}
\def\pDensUnder{\underline{\mathfrak{p}}}
\def\diagonal{\mbox{diagonal}}
\def\nFull{n_{\mbox{{\tiny full}}}}
\def\etaTrue{\eta_{\mbox{{\tiny true}}}}
\def\etaTrueOne{\eta_{\mbox{{\tiny true}},1}}
\def\etaTrueTwo{\eta_{\mbox{{\tiny true}},2}}
\def\kappaTrue{\kappa_{\mbox{{\tiny true}}}}
\def\bcnew{\bc_{\mbox{{\tiny new}}}}

\def\ssigma{s_{\sigma}}
\def\bMtilde{{\widetilde \bM}}
\def\vecof{\mbox{\rm vec}}
\def\lambdaJJ{\lambda_{\mbox{\tiny JJ}}}

\def\pDens{\mathfrak{p}}
\def\qDens{\mathfrak{q}}
\def\logML{\underline{\ell}}

\def\PolyaGamma{\mbox{P\'olya-Gamma}}
\def\NegBinom{\mbox{Negative-Binomial}}
\def\sigmabeta{\sigma_{\beta}}
\def\calphakappanew{c(\alpha|\kappa)_{\mbox{\tiny new}}}
\def\pDensPG{\pDens_{\mbox{\tiny PG}}}
\def\sigsqbeta{\sigma_{\beta}^2}
\def\bsigsq{\bsigma^2}
\def\bbetabuVec{\left[
\begin{array}{c}   
\bbeta\\
\bu
\end{array}
\right]}
\def\yTone{\by^T\bone}
\def\CTone{\bC^T\bone}
\def\CTy{\bC^T\by}
\def\myUniversita{Universit$\grave{\mbox{a}}$}
\def\nwarm{n_{\mbox{\tiny warm}}}
\def\bywarm{\by_{\mbox{\scriptsize warm}}}
\def\bCwarm{\bC_{\mbox{\scriptsize warm}}}

\def\qwarm{\qDens_{\mbox{\tiny warm}}}
\def\muwarm{\mu_{\mbox{\tiny warm}}}
\def\sigmawarm{\sigma_{\mbox{\tiny warm}}}
\begin{document}

\ifthenelse{\boolean{DoubleSpaced}}{\doublespacing}{}

\vskip5mm
\centerline{\Large\bf Variational Inference for Count Response}
\vskip2mm
\centerline{\Large\bf Semiparametric Regression: A Convex Solution}
\vskip5mm
\centerline{\normalsize\sc Virginia Murru and Matt P. Wand}
\vskip5mm
\centerline{\textit{\myUniversita\ di Padova and University of Technology Sydney}}
\vskip5mm
\centerline{23rd February 2026}
\vskip5mm
\centerline{\large\bf Abstract}
\vskip2mm


We develop a version of variational inference for Bayesian count
response regression-type models that possesses attractive 
attributes such as convexity and closed form updates.
The convex solution aspect entails numerically stable
fitting algorithms, whilst the closed form aspect makes
the methodology fast and easy to implement. The essence
of the approach is the use of P\'olya-Gamma augmentation
of a Negative Binomial likelihood, a finite-valued prior
on the shape parameter and the structured mean field 
variational Bayes paradigm. The approach applies to general
count response situations. For concreteness, we focus on
generalized linear mixed models within the semiparametric
regression class of models. Real-time fitting is also described.


\section{Introduction}\label{sec:intro}

Variational approximation is an alternative to Monte Carlo methods
for Bayesian inference and can be useful in applications where
speed and scalability are at a premium. For count response semiparametric
regression, Luts \myand Wand (2015) provided variational inference
algorithms using a fixed form, or semiparametric, 
mean field variational Bayes approach. Despite its good accuracy,
the non-convexity of fixed form approaches can lead to 
numerical problems. For example, in the simulation study described
in Section 3 of Wand \myand Yu (2022), the fixed form
variational approach to Poisson nonparametric regression failed
to converge properly in 13.6\% of the replications. In this
article we devise a variational inference approach for which
all component optimization problems are convex. The Luts \myand Wand (2015)
methodology also involved some numerical integration steps, whereas
our new approach has totally closed form updates. These attributes
lead to fast and stable algorithms that are easy to implement.

Our approach to variational inference for count response semiparametric
regression has similarities with the binary response model approaches
of Jaakola \myand Jordan (2000) and Durante \myand Rigon (2019).
For a given model the coordinate ascent updates in these two articles 
are identical, but the latter makes use of P\'olya-Gamma
augmentation (e.g. Polson \textit{et al.}, 2013) which has
the advantage of couching the algorithm within the ordinary
mean field variational Bayes framework. As explained in, for example,
Pillow \myand Scott (2012) and Zhou \textit{et al.} (2012) P\'olya-Gamma
augmentation can also aid the fitting of Bayesian regression-type
models with Negative Binomial likelihoods. An additional 
difficulty compared with the binary response setting is
approximate Bayesian inference for the Negative Binomial
shape parameter, which we denote here by $\kappa$. We deal
with this problem via a \emph{structured} mean field variational
Bayes approach (e.g.\ Saul \myand Jordan, 1996; 
Wand \textit{et al.} 2011). This involves
restriction of $\kappa$ to a finite set and performing
a variational version of Bayesian model averaging, where
individual models correspond to the atoms of the prior
distribution of $\kappa$. Since $\kappa$ is more of
a nuisance parameter, and any finite set can be specified,
there is little cost to this discretisation of $\kappa$.

Apart from Luts \myand Wand (2015), we are aware of some other
approaches to variational inference for count response regression-type
models. In particular, Zhou \textit{et al.} (2012) and Miao \textit{et al.}
(2020) each contain variational inference algorithms which are also 
based on P\'olya-Gamma augmentation of Negative Binomial likelihoods.
Their approaches make use of Logarithmic series or, equivalently,
Chinese Restaurant Process, representations of Negative Binomial
response models. However, when this representation is combined
with P\'olya-Gamma augmentation there is no single joint distribution
to which minimum Kullback-Leibler divergence, the underpinning
of mean field variational Bayes, is being applied. In addition,
the square-root quantity that arises in the tilting parameter
of the P\'olya-Gamma $\qDens$-density, such as the $\xi_i^{(t)}$
quantity in Algorithm 2 of Durante \myand Rigon (2019) and the 
$\bc_{\qDens(\balpha|\kappa)}$ quantity in Algorithm \ref{alg:BCHalgo}
of this article, is absent from the Zhou \textit{et al.} (2012) 
and Miao \textit{et al.} (2020) algorithms. Our implementations
of the Miao \textit{et al.} (2020) approach resulted in
low accuracy compared with the variational approximation strategy 
developed here.

Our new convex solution for count response semiparametric
regression also benefits real-time fitting and inference. 
In Section 5 and Algorithm 2 of Luts \myand Wand (2015) we
presented an online variational algorithm for real-time
count response semiparametric regression. However, 
storage of the predictor data and spline basis design
matrices was required. In Section \ref{sec:realtime} of
this article we present a new real-time algorithm 
for the same class of models that is \emph{purely}
online, in that only sufficient statistics-type quantities
need to be updated and stored. The streaming data can be 
discarded after they are processed.

Notation used throughout this article
is given in Section \ref{sec:notation}. 
Section \ref{sec:modDesc} describes the specific
count response semiparametric regression
models to which we gear our methodological
development. The new variational inference
scheme is described in Section \ref{sec:VIscheme}.
Real-time semiparametric using online adaptations
of our variational inference approach is described
in Section \ref{sec:realtime}. Section \ref{sec:numRes}
contains numerical results. We provide some
conclusions in Section \ref{sec:conclusion}.

\subsection{Notation}\label{sec:notation}

A real-valued function, which is defined and prominent in 
Jaakkola \myand Jordan (2000), and also important here is
that given by
\begin{equation}
\lambdaJJ(x)\equiv\frac{\tanh(x/2)}{4x},\quad x\in\real.
\label{eq:lamJJdefn}
\end{equation}
Scalar functions applied to a vector are evaluated
in an element-wise fashion. For example, 
$\cosh([3\ 11]^T)\equiv[\cosh(3)\ \cosh(11)]^T$.
Similarly, if $\bv$ is a column vector then 
$\bv^2$ is the vector of element-wise squares
and $\Vert\bv\Vert\equiv\sqrt{\bv^T\bv}$ is
the Euclidean norm of $\bv$. 
The notation $\diag(\bv)$ is used for the diagonal
matrix containing the entries of $\bv$ along its
diagonal. If $\bM$ is a $d\times d$ square matrix then 
$\mbox{diagonal}(\bM)$ is the $d\times 1$ vector
containing the diagonal entries of $\bM$. 
Also, $\bone$ is a column vector of ones.
The symbol $\simind$ is shorthand for
``independently distributed as''. 

\section{Model Description}\label{sec:modDesc}

Throughout this article we focus on the following
count response Bayesian semiparametric regression model:
\begin{equation}
\begin{array}{c}
y_i|\,\bbeta,\bu,\kappa \simind
\NegBinom\big(\,\exp\{(\bX\bbeta+\bZ\bu)_i\},\kappa\big),
\quad 1 \leq i \leq n,\\[2ex]
\alpha_i|\,y_i,\bbeta,\bu,\kappa \simind
\PolyaGamma\big(y_i+\kappa, (\bX\bbeta+\bZ\bu)_i-\log(\kappa)\big),\\[3ex]
\bu|\,\sigma_{1}^2,
\ldots,\sigma_{r}^2\sim N(\bzero,\mbox{blockdiag}
(\sigma_{1}^2\,\bI_{K_1},\ldots,\sigma_{r}^2\,\bI_{K_r})),
\quad
\bbeta\sim N(\bzero,\sigsqbeta\bI_p),
\\[2ex]
\quad
\sigma_j\simind\mbox{Half-Cauchy}(\ssigma),\quad 1\le j\le r,\\[2ex]
\mbox{and}\ \kappa\ \mbox{has a discrete prior with atoms $\Ksc$ and
probabilities $\pDens(\kappa)$, $\kappa\in\Ksc$.}
\end{array}
\label{eq:negBinModel}
\end{equation}
In (\ref{eq:negBinModel}) $\sigma_{\beta}>0$ and $\ssigma>0$ are
user-specified hyperparameters. Distributional notation used
in (\ref{eq:negBinModel}) is defined later in this section.

Model (\ref{eq:negBinModel}) is a variant of what Zhao \textit{et al.} (2006)
label \emph{general design} generalized linear mixed models. 
Here, $\bX$ is a design matrix attached to
the unpenalized vector of coefficients $\bbeta$ and  
$\bZ$ is a design matrix attached to the penalized vector 
of coefficients $\bu$, which is partitioned into $r$ sub-vectors 
of sizes $K_1,\ldots,K_r$.
The $\sigma_j$, $1\le j\le r$, control the amount of penalization 
for the coefficients within the $j$th sub-vector.
As explained in Section 2 of Zhao \textit{et al.} (2006), 
the $\bX\bbeta+\bZ\bu$ component of (\ref{eq:negBinModel})
is a very versatile structure and special 
cases of (\ref{eq:negBinModel})
include nested random effects models for grouped (e.g.\ longitudinal)
data, crossed random effects models for item response data,
generalized additive models, generalized additive mixed models,
varying-coefficients models and low-ranking kriging.
Specific examples of the $\bX$ and $\bZ$ design matrices, along with
$\bu$ vector partitioning, are given in Section \ref{sec:numRes}.

\begin{figure}[!h]
\centering
\includegraphics[width=0.6\textwidth]{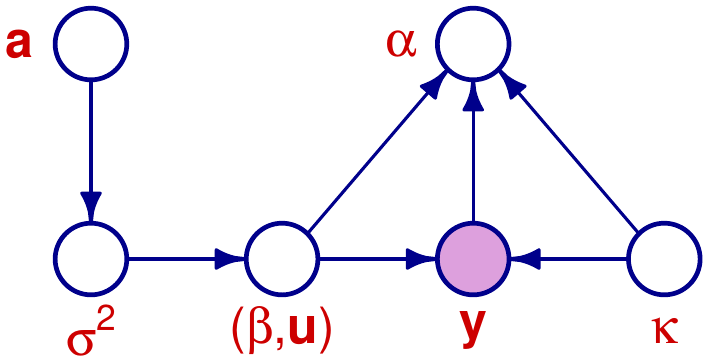}
\caption{\textit{
Directed acyclic graph representation of model (\ref{eq:negBinModel})
with incorporation of the $\ba=(a_1,\ldots,a_r)$ auxiliary
variables as in (\ref{eq:HCtoIG}). The $\by$ node is shaded to 
indicate that it contains observed data.
}}
\label{fig:nbcDAG}
\end{figure}

The distributional notation in (\ref{eq:negBinModel}) is such that
$x\sim\mbox{Negative-Binomial}(\mu,\kappa)$ denotes that the random
variable $x$ has Negative Binomial distribution with mean
$\mu>0$ and shape parameter $\kappa>0$ with
probability mass function:
$$
\pDens(x)=
\frac{\kappa^{\kappa}\Gamma(x+\kappa)\mu^{x}}
{\Gamma(\kappa)\big(\kappa+\mu\big)^{x+\kappa}\Gamma(x+1)},
\quad x=0,1,2,\ldots.
$$
The $\PolyaGamma$ distributional notation matches that
used in Polson \textit{et al.} (2013) and is described
in Section \ref{sec:PGdistn} of the online supplement.
Also, $x\sim\mbox{Half-Cauchy}(s)$, with scale parameter $s>0$, 
means that the random variable $x$ has density function
$$\pDens(x)=2/[\pi\{1+(x/s)^2\}s],\quad x>0.$$
Variational inferential tractability is aided by the replacement of 
$\sigma_j\sim\mbox{Half-Cauchy}(\ssigma)$ by
\begin{equation}
\sigma^2_j|a_j
\sim\mbox{Inverse-Gamma}(\smhalf,1/a_j),\ \ 
a_j\sim\mbox{Inverse-Gamma}(\smhalf,1/\ssigma^2),
\label{eq:HCtoIG}
\end{equation}
where $x\sim\mbox{Inverse-Gamma}(\xi,\lambda)$
means that $x$ has density function 
$$\pDens(x)=\frac{\lambda^{\xi}}{\Gamma(\xi)}\,x^{-\xi-1}
\exp\big(-\lambda/x\big)
,\quad x>0.
$$
If we let 
$$\by\equiv(y_1,\ldots,y_n),\quad
\balpha\equiv(\alpha_1,\ldots,\alpha_n),\quad
\bsigsq\equiv(\sigma_1^2,\ldots,\sigma_r^2)\quad
\mbox{and}\quad
\ba\equiv(a_1,\ldots,a_r)
$$
then model (\ref{eq:negBinModel}) has directed acyclic graph
representation as shown in Figure \ref{fig:nbcDAG}.

Model (\ref{eq:negBinModel}) is similar to the class
of Negative Binomial response models considered by Luts \myand Wand (2015). 
The differences are the presence of the P\'olya-Gamma auxiliary variables
and the imposition of a discrete prior on the shape
parameter $\kappa$.

\subsection{Extension to Covariance Matrix Parameters for Random Effects}

To simplify the exposition, model (\ref{eq:negBinModel}) only contains
scalar variance parameters. Extensions to covariance matrix
parameters arise in the case of random intercept and slope models. 
A simple example of such a model having count responses is,
for  $1\le i\le m$ and $1\le j\le n_i$,
$$
y_{ij}|\beta_0,\beta_1,u_{0i},u_{1i}\simind
\mbox{Negative-Binomial}
\Big(\exp\big(\beta_0+u_{0i}+(\beta_1+u_{1i}\,x_{ij})\big),\kappa\Big)
$$
where
$$
\left[
\begin{array}{c}
u_{0i}\\
u_{1i} \\
\end{array}
\right]\Big|\bSigma\simind N(\bzero,\bSigma)
$$
and $\bSigma$ is an unstructured $2\times 2$ covariance matrix.
Our new variational methodology has a straightforward
extension to models containing covariance matrix parameters.

\section{Variational Inference Scheme}\label{sec:VIscheme}

Consider the Negative Binomial response additive model model
given by (\ref{eq:negBinModel}) with the auxiliary 
variable replacement (\ref{eq:HCtoIG}) and directed
acyclic graph representation given in Figure \ref{fig:nbcDAG}.
At its most general level, variational approximation of the full joint
posterior density function of the parameters in (\ref{eq:negBinModel}) 
involves
\begin{equation}
\pDens(\bbeta,\bu,\kappa,\balpha,\bsigsq,\ba|\by)
\approx\qDens(\bbeta,\bu,\kappa,\balpha,\bsigsq,\ba)
\label{eq:JackHorner}
\end{equation}
where the $\qDens$-density on the right-hand side of 
(\ref{eq:JackHorner}) is subject to specific restrictions.
An initial restriction to consider is one involving the
following product density form:
\begin{equation}
\qDens(\bbeta,\bu,\kappa,\balpha,\bsigsq,\ba)
=\qDens(\bbeta,\bu,\ba)\qDens(\bsigsq,\balpha)\qDens(\kappa).
\label{eq:KevsWhistle}
\end{equation}
Such a restriction is an example of ordinary mean field variational
Bayes (e.g.\ Wainwright \myand Jordan, 2008) and, assuming 
tractability, the optimal $\qDens$-densities can be 
found via coordinate ascent (e.g.\ Algorithm 1 of Ormerod \myand Wand, 2010)
based on consistency conditions such as 
{\setlength\arraycolsep{1pt}
\begin{eqnarray*}
\qDens^*(\bbeta,\bu,\ba)&\propto&\exp\Big(E_{\qDens(-(\bbeta,\bu,\ba))}
\big[\log\{\pDens(\bbeta,\bu,\ba|\rest)\}\big]\Big)
\quad\mbox{and}\\[0ex]
\qDens^*(\kappa)&\propto&\exp\Big(E_{\qDens(-\kappa)}
\big[\log\{\pDens(\kappa|\rest)\}\big]\Big).
\end{eqnarray*}
}
Here, for example, $E_{\qDens(-(\bbeta,\bu,\ba))}$ denotes expectation
with respect to all $\qDens$-densities other than $(\bbeta,\bu,\ba)$.
Also, $\pDens(\bbeta,\bu,\ba|\rest)$ denotes the conditional density
function of $(\bbeta,\bu,\ba)$ given the rest of the random variables
in the model and called the full conditional density function
of $(\bbeta,\bu,\ba)$. Attributes of (\ref{eq:negBinModel}) such as   
P\'olya-Gamma augmentation lead to closed formed expressions 
for all but one of the \emph{full conditional} density functions. The exception
is $\pDens(\kappa|\rest)$ which, as shown in Section \ref{sec;ordMFVB} 
of the online supplement, does not admit a closed form expression. 
This makes ordinary mean field variational Bayes impractical 
for model (\ref{eq:negBinModel}).

As a way of overcoming the difficulties with ordinary
mean field variational Bayes we note that
\begin{equation}
\pDens(\bbeta,\bu,\kappa,\balpha,\bsigsq,\ba|\by)=
\pDens(\bbeta,\bu,\balpha,\bsigsq,\ba|\kappa,\by)\pDens(\kappa|\by)
\label{eq:AltraniBook}
\end{equation}
and then consider approximation of the right-hand
side of (\ref{eq:AltraniBook}) using the product density form:
\begin{equation}
\qDens(\bbeta,\bu,\kappa,\balpha,\bsigsq,\ba)
=\qDens(\bbeta,\bu,\ba|\kappa)\qDens(\bsigsq,\balpha|\kappa)\qDens(\kappa).
\label{eq:theProdDensRestric}
\end{equation}
which differs from (\ref{eq:KevsWhistle}) due to its
conditioning on $\kappa$. Noting that $\kappa$ is confined to a finite set, we apply a
\emph{structured} mean field variational Bayes approach to
obtaining the optimal $\qDens$-densities under restriction
(\ref{eq:theProdDensRestric}). In essence, this involves performing
ordinary mean field variational Bayes for each $\kappa\in\Ksc$
as if $\kappa$ is fixed and then obtaining a weighted 
average of the resultant $\qDens$-densities that depends
on the $\kappa$ prior distribution and the variational-approximate 
marginal log-likelihoods. A summary of this approach is given
in Section 3.1 of Wand \textit{et al.} (2011). From Section 10.1.1 of Bishop (2006), 
each individual (over $\kappa\in\Ksc$) 
mean field optimization problem is convex. 
Also, induced factorization theory (e.g.\ Bishop, 2006; Section 10.2.5)
implies that we have the additional product density
forms $\qDens(\bbeta,\bu,\ba|\kappa)=\qDens(\bbeta,\bu|\kappa)\qDens(\ba|\kappa)$
and $\qDens(\bsigsq,\balpha|\kappa)=\qDens(\bsigsq|\kappa)\qDens(\balpha|\kappa)$
even though these are not imposed at the outset.

\begin{algorithm}[!th]
\begin{center}
\begin{minipage}[t]{160mm}
\hrule
\begin{small}
\begin{itemize}
\item[] Inputs: $\by$ $(n\times 1)$,
\ response vector having all entries non-negative integers,
\item[] $\qquad\quad\ \bC$\ $\left(n\times\left(p+\sum_{j=1}^rK_j\right)\right)$, 
\ \ combined design matrix,
\item[]$\qquad\quad\ \sigma_{\beta},\ssigma>0$, hyperparameters for the 
$\bbeta$ and $\sigma_j$ prior distributions
\item[]$\qquad\quad\ \Ksc$, a finite set of positive numbers 
corresponding to the atoms of the prior
\item[]$\qquad\quad\ $ distribution of $\kappa$.
\item[] Initialize: $\mu_{q(1/\sigma_j^2|\kappa)}>0$, $1\le j\le r$,
and
$\bc_{\qDens(\balpha|\kappa)}$ ($n\times1$) all entries positive,
for each $\kappa\in\Ksc$.
\item[] For $\kappa\in\Ksc$:
\begin{itemize}
\item[] Cycle:
\begin{itemize}
\item[] $\bmu_{\qDens(\balpha|\kappa)}
\longleftarrow 2(\by+\kappa\bone)\odot\lambdaJJ(\bc_{\qDens(\balpha|\kappa)})$
\item[] $\bM_{q(1/\bsigma^2|\kappa)} \longleftarrow 
\mbox{blockdiag}(\sigma_{\beta}^{-2}\bI_p,\mu_{q(1/\sigma_1^2|\kappa)}
\bI_{K_1},\ldots,\mu_{q(1/\sigma_r^2|\kappa)}\bI_{K_r})$
\item[] $\bSigma_{\qDens(\bbeta,\bu|\kappa)}
\longleftarrow \Big\{\bC^T\diag\big(\bmu_{\qDens(\balpha|\kappa)}\big)\bC
+\bM_{\qDens(1/\bsigsq|\kappa)} \Big\}^{-1}$
\item[] $\bmu_{\qDens(\bbeta,\bu|\kappa)}\longleftarrow 
\bSigma_{\qDens(\bbeta,\bu|\kappa)}
\Big\{\smhalf(\CTy-\kappa\CTone)+\log(\kappa)
\bC^T\bmu_{\qDens(\balpha|\kappa)}\Big\}
$
\item[] $\bc_{\qDens(\balpha|\kappa)}\longleftarrow\sqrt{\mbox{diagonal
}\Big(\bC\bSigma_{\qDens(\bbeta,\bu|\kappa)}\bC^T\Big)
+\big(\bC\bmu_{\qDens(\bbeta,\bu|\kappa)}-\log(\kappa)\bone\big)^2}$
\item[] For $j=1,\ldots,r:$
\begin{itemize}
\item[] $\lambda_{\qDens(a_j|\kappa)}\longleftarrow
\mu_{\qDens(1/\sigma_j^2|\kappa)}+\ssigma^{-2}$
\ \ \ ;\ \ \ 
$\mu_{\qDens(1/a_j|\kappa)}\longleftarrow 
1/\lambda_{\qDens(a_j|\kappa)}$
\item[] $\lambda_{\qDens(\sigma^2_j|\kappa)}\longleftarrow
\mu_{\qDens(1/a_j|\kappa)}
+\smhalf
\big\{
\Vert\bmu_{\qDens(\bu_j|\kappa)}\Vert^2+
\mbox{tr}(\bSigma_{\qDens(\bu_j|\kappa)})\big\}
$
\item[]
$\mu_{\qDens(1/\sigma_j^2|\kappa)}\longleftarrow 
(K_j+1)/\big(2\lambda_{\qDens(\sigma^2_j|\kappa)}\big)$
\end{itemize}
\item[] $\logML_{\qDens}(\kappa)\longleftarrow
\smhalf\bmu_{\qDens(\bbeta,\bu|\kappa)}^T\big(\CTy-\kappa\,\CTone\big)
-(\by+\kappa\bone)^T\log\left\{
\cosh\Big(\smhalf c_{\qDens(\balpha|\kappa)}\Big)\right\}$
\item[] $\qquad\qquad-\displaystyle{
\frac{\Vert\bmu_{\qDens(\bbeta|\kappa)}\Vert^2
+\tr(\bSigma_{\qDens(\bbeta|\kappa)})}{2\sigsqbeta}}
+\smhalf\log|\bSigma_{\qDens(\bbeta,\bu|\kappa)}|$ 
\item[] For $j = 1,\ldots,r:$
\begin{itemize}
\item[] $\logML_{\qDens}(\kappa)\longleftarrow
\logML_{\qDens}(\kappa)+\mu_{\qDens(1/\sigma_j^2|\kappa)}
\big\{\lambda_{\qDens(\sigma^2_j|\kappa)}
-\mu_{\qDens(1/a_j|\kappa)}
-\smhalf\Vert\bmu_{\qDens(\bu_j|\kappa)}\Vert^2
-\smhalf\tr\big(\bSigma_{\qDens(\bu_j|\kappa)}\big)\big\}$
\item[] $\qquad\qquad\quad+\mu_{\qDens(1/a_j|\kappa)}
\big(\lambda_{\qDens(a_j|\kappa)}-\ssigma^{-2}\big)
-\smhalf(K_j+1)\log\big(\lambda_{\qDens(\sigma^2_j|\kappa)}\big)
-\log\big(\lambda_{\qDens(a_j|\kappa)}\big)$
\end{itemize}
\end{itemize}
\item[] until the increase in $\logML_{\qDens}(\kappa)$ is negligible.
\item[] $\logML(\kappa)\longleftarrow\logML_{\qDens}(\kappa)+\bone^T\log\{\Gamma(\by+\kappa\bone)\}
+n\big[\smhalf\kappa\log(\kappa)-\log(2)\kappa-\log\{\Gamma(\kappa)\}\big]
-\smhalf\log(\kappa)(\yTone)$
\end{itemize}
\item[] Outputs: $\big\{\bmu_{\qDens(\bbeta,\bu|\kappa)},  
\bSigma_{\qDens(\bbeta,\bu|\kappa)},
\lambda_{\qDens(\sigma^2_j|\kappa)},\logML(\kappa):\kappa\in\Ksc,\ 1\le j\le r\big\}$
\end{itemize}
\end{small}
\hrule
\end{minipage}
\end{center}
\caption{\textit{Structured mean field variational Bayes algorithm
for achieving approximate Bayesian inference for 
model (\ref{eq:negBinModel}) according to product density
restriction (\ref{eq:theProdDensRestric}).}}
\label{alg:BCHalgo}
\end{algorithm}
%

Algorithm \ref{alg:BCHalgo} describes a suite of coordinate
descent algorithms for obtaining, iteratively, the parameters of
the optimal $\qDens$-densities. It uses the following 
definitions:
$$\bC\equiv[\bX\ \bZ]$$
and
\begin{equation}
\bu_j\ \mbox{is the $K_j\times 1$ block of $\bu$
according to the partition}\ 
\bu=[\bu_1^T\ \cdots\ \bu_r^T]^T.
\label{eq:uparticDefn}
\end{equation}
It also involves the $\lambdaJJ$ function defined by (\ref{eq:lamJJdefn}).


Based on the output from Algorithm \ref{alg:BCHalgo}
and the structured mean field variational Bayes formulae given 
in Section 3.1 of Wand \textit{et al.} (2011),
the approximate posterior distributions of the model
parameters are obtained as follows:
$$
\qDens^*(\kappa)=\frac{\pDens(\kappa)\logML(\kappa)}
{{\displaystyle\sum_{\kappa'\in\Ksc}}\pDens(\kappa')\logML(\kappa')},\quad
\kappa\in\Ksc,
$$
\begin{eqnarray*}
&&\qDens^*(\bbeta,\bu)=\sum_{\kappa\in\Ksc}\qDens^*(\kappa)
\big|2\pi\bSigma_{\qDens(\bbeta,\bu|\kappa)}\big|^{-1/2}\\[1ex]
&&\qquad\qquad\quad\times\exp\left\{-\smhalf\left(
\bbetabuVec-\bmu_{\qDens(\bbeta,\bu|\kappa)}\right)^T 
\bSigma_{\qDens(\bbeta,\bu|\kappa)}^{-1}    
\left(\bbetabuVec-\bmu_{\qDens(\bbeta,\bu|\kappa)}\right)
\right\}
\end{eqnarray*}
and
$$\qDens^*(\sigma_j^2)
=\sum_{\kappa\in\Ksc}
\left\{
\frac{\qDens^*(\kappa)\lambda_{\qDens(\sigma^2_j|\kappa)}^{(K_j+1)/2}}
{\Gamma\big(\smhalf(K_j+1)\big)}\right\}
(\sigma_j^2)^{-(K_j+1)/2-1}
\exp\left(-\frac{\lambda_{\qDens(\sigma^2_j|\kappa)}}{\sigma_j^2}\right),
\ \ \sigma_j^2>0,\ \  1\le j\le r.
$$

\subsection{Streamlined Variational Inference Alternatives}

In grouped data situations, the sub-blocks of the $\bZ$ matrix
corresponding to random effects typically are quite sparse. 
Algorithm \ref{alg:BCHalgo} is still valid for such $\bZ$
matrices but, for large numbers of groups, tends to be
inefficient when applied na\"{\i}vely. Instead streamlined variational 
inference alternatives, which take advantage of sparse structure
in $\bZ$ matrices, are recommended. Lee \myand Wand (2016)
provides streamlined variational methodology for 
binary response semiparametric regression models for
grouped data. The same ideas apply to the count response
setting treated here. The details will appear in the
first author's upcoming doctoral dissertation.

\section{Real-Time Count Response Semiparametric Regression}\label{sec:realtime}

The structured mean field variational Bayes approach used in
Algorithm \ref{alg:BCHalgo} also lends itself to online fitting
of streaming data. This allows real-time count response semiparametric
regression. Moreover, unlike in Luts \myand Wand (2015), this new approach
has the attractive aspects of only requiring
low-dimensional sufficient statistics quantities to be stored
and updated. There is no need to keep the full data in memory.

%
\begin{algorithm}[!th]
\begin{center}
\begin{minipage}[t]{160mm}
\hrule
\begin{small}
\begin{itemize}
\item[1.] Perform batch-based tuning runs analogous to those
described in Algorithm 2' of  Luts, Broderick \myand Wand (2014)
and determine a warm-up sample size $\nwarm$ for which 
convergence is validated.
\item[2.] Set $\bywarm$ and $\bCwarm$ 
to be the response vector
and design matrix and, for each $\kappa\in\Ksc$, 
let $\bc_{\qDens(\balpha|\kappa)}$ be the vector
of P\'olya-Gamma variational tilting parameters,
based on the first $\nwarm$ observations. 
Then set $n\leftarrow\nwarm$, $\by^T\bone\leftarrow\bywarm^T\bone$,
$\bC^T\bone\leftarrow\bCwarm^T\bone$,
$\bC^T\by\leftarrow\bCwarm^T\bywarm$.
For each $\kappa\in\Ksc$ set 
$\bone^T\log\{\Gamma(\by+\kappa\bone)\}\longleftarrow
\bone^T\log\{\Gamma(\bywarm+\kappa\bone)\}$,
$\bC^T\lambdaJJ(\bc_{\qDens(\balpha|\kappa)})
\longleftarrow \bCwarm^T\lambdaJJ(\bc_{\qDens(\balpha|\kappa)})$
and similar sufficient statistics quantities that appear in Step 3 below.
Also, set
$\bmu_{\qDens(\bbeta,\bu|\kappa)},
\ \bSigma_{\qDens(\bbeta,\bu|\kappa)},
\ \mu_{q(1/\sigma^2_{u1}|\kappa)},\ldots,\mu_{q(1/\sigma^2_{ur}|\kappa)}$
to be the values for these quantities obtained in the batch-based 
tuning run with sample size $\nwarm$.
\item[3.] Let $\qwarm^*(\kappa)$, $\kappa\in\Ksc$, be the $\qDens$-density
of $\kappa$ obtained from feeding  $\bywarm$, $\bCwarm$, $\sigma_{\beta}$
and $\ssigma$ into Algorithm \ref{alg:BCHalgo}.
\item[4.] Cycle:
\begin{itemize}
\item[] read in $\ynew$ and $\bcnew$\ \ \ ;\ \ \ $n\longleftarrow n+1$
\item[] $\by^T\bone\longleftarrow\by^T\bone+\ynew$\ \ \ ;\ \ \ 
$\bC^T\bone\longleftarrow\bC^T\bone+\bcnew$\ \ \ ;\ \ \ 
$\bC^T\by\longleftarrow\bC^T\by+\bcnew\ynew$
\item[] Obtain the atom set $\Ksc_n\subseteq\Ksc$ for the current sample size
such that the retained\\
atoms coax $\qDens^*(\kappa)$ to be more concentrated around 
the posterior mean of $\qwarm^*(\kappa)$.\\
(A practical recommendation for this step is described in 
Section \ref{sec:atomsRedu}.)
\item[] For $\kappa\in\Ksc_n$:
\begin{itemize}
\item[] $\bone^T\log\{\Gamma(\by+\kappa\bone)\}\longleftarrow
\bone^T\log\{\Gamma(\by+\kappa\bone)\}+\log\{\Gamma(\ynew+\kappa)\}$
\item[] $\calphakappanew\longleftarrow\sqrt{\bcnew^T
\bSigma_{\qDens(\bbeta,\bu|\kappa)}\bcnew+
\{\bcnew^T\bmu_{\qDens(\bbeta,\bu|\kappa)}-\log(\kappa)\}^2}$
\item[] $\bC^T\lambdaJJ(\bc_{\qDens(\balpha|\kappa)})
\longleftarrow \bC^T\lambdaJJ(\bc_{\qDens(\balpha|\kappa)})
+\lambdaJJ(\calphakappanew)\bcnew$
\item[] $\bC^T\big(\by\odot\lambdaJJ(\bc_{\qDens(\balpha|\kappa)})\big)
\longleftarrow \bC^T\big(\by\odot\lambdaJJ(\bc_{\qDens(\balpha|\kappa)})\big)
+\ynew\lambdaJJ(\calphakappanew)\bcnew$
\item[] $\bC^T\diag\big(\lambdaJJ(\bc_{\qDens(\balpha|\kappa)})\big)\bC
\longleftarrow \bC^T\diag\big(\lambdaJJ(\bc_{\qDens(\balpha|\kappa)})\big)\bC
+\lambdaJJ(\calphakappanew)\bcnew\bcnew^T$
\item[] $\bC^T\diag\big(\by\odot\lambdaJJ(\bc_{\qDens(\balpha|\kappa)})\big)\bC
\longleftarrow \bC^T\diag\big(\by\odot\lambdaJJ(\bc_{\qDens(\balpha|\kappa)})\big)\bC$
\item[]$\qquad\qquad\qquad\qquad\qquad\qquad\qquad\qquad
+\ynew\lambdaJJ(\calphakappanew)\bcnew\bcnew^T$
\item[] $\bone^T\log\{\cosh(\smhalf\bc_{\qDens(\balpha|\kappa)})\}
\longleftarrow\bone^T\log\{\cosh(\smhalf\bc_{\qDens(\balpha|\kappa)})\}
+\log\{\cosh(\smhalf\calphakappanew)\}
$
\item[] $\by^T\log\{\cosh(\smhalf\bc_{\qDens(\balpha|\kappa)})\}
\longleftarrow\by^T\log\{\cosh(\smhalf\bc_{\qDens(\balpha|\kappa)})\}$
\item[] $\qquad\qquad\qquad\qquad\qquad\qquad\qquad\qquad
+\ynew\log\{\cosh(\smhalf\calphakappanew)\}
$
\item[] $\bM_{q(1/\bsigma^2|\kappa)} \longleftarrow 
\mbox{blockdiag}(\sigma_{\beta}^{-2}\bI_p,\mu_{q(1/\sigma_1^2|\kappa)}
\bI_{K_1},\ldots,\mu_{q(1/\sigma_r^2|\kappa)}\bI_{K_r})$
\item[] $\bSigma_{\qDens(\bbeta,\bu|\kappa)}\longleftarrow \Big\{
2\bC^T\diag\big(\by\odot\lambdaJJ(\bc_{\qDens(\balpha|\kappa)})\big)\bC$
\item[] $\null\qquad\qquad\qquad\qquad\qquad
+2\kappa\bC^T\diag\big(\lambdaJJ(\bc_{\qDens(\balpha|\kappa)})\big)\bC
+\bM_{q(1/\bsigma^2|\kappa)}\Big\}^{-1}$
\item[] $\bmu_{\qDens(\bbeta,\bu|\kappa)}\longleftarrow \bSigma_{\qDens(\bbeta,\bu|\kappa)}
\Big(\smhalf(\bC^T\by-\kappa\bC^T\bone)$
\item[]
$\null\qquad\qquad\qquad
+2\log(\kappa)\big\{\bC^T\big(\by\odot\lambdaJJ(\bc_{\qDens(\balpha|\kappa)})\big)
+\kappa \bC^T\lambdaJJ(\bc_{\qDens(\balpha|\kappa)})\big\}\Big)$
\item[] For $j = 1,\ldots,r:$
\begin{itemize}
\item[] $\lambda_{\qDens(a_j|\kappa)}\longleftarrow
\mu_{\qDens(1/\sigma_j^2|\kappa)}+\ssigma^{-2}$
\ \ \ ;\ \ \ 
$\mu_{\qDens(1/a_j|\kappa)}\longleftarrow 
1/\lambda_{\qDens(a_j|\kappa)}$
\item[] $\lambda_{\qDens(\sigma^2_j|\kappa)}\longleftarrow
\mu_{\qDens(1/a_j|\kappa)}
+\smhalf
\big\{
\Vert\bmu_{\qDens(\bu_j|\kappa)}\Vert^2+
\mbox{tr}(\bSigma_{\qDens(\bu_j|\kappa)})\big\}
$
\item[]
$\mu_{\qDens(1/\sigma_j^2|\kappa)}\longleftarrow 
(K_j+1)/\big(2\lambda_{\qDens(\sigma^2_j|\kappa)}\big)$
\end{itemize}
\end{itemize}
\item[]\textit{continued on a subsequent page $\ldots$}
\end{itemize}
\end{itemize}
\end{small}
\hrule
\end{minipage}
\end{center}
\caption{\textit{Online structured mean field variational Bayes algorithm
for achieving real-times approximate Bayesian inference for 
model (\ref{eq:negBinModel}) according to product density
restriction (\ref{eq:theProdDensRestric}).}}
\label{alg:ONLalgo}
\end{algorithm}
%

Our proposed algorithm for real-time count response semiparametric
regression is similar to that used in Section 3 of
Luts \textit{et al.} (2014) for real-time
binary response semiparametric regression. The main difference
is the presence of the $\kappa$ parameter and its finite set
restriction. The essence of the approach is to express each
of $\qDens$-density parameters in terms of sufficient statistics
quantities such as $\bC^T\by$. If the current sample size is 
$n$ then 
$$\bC^T\by=\sum_{i=1}^n \bc_iy_i\quad\mbox{where}
\quad\bc_i\equiv\mbox{$i$th row of $\bC$}.$$
When a new observation arrives with response value $\ynew$ and
corresponding design matrix row $\bcnew$ then the
$$\bC^T\by\quad\mbox{sufficient statistic is incremented by}\quad \bcnew\ynew.$$
Similarly, for each $\kappa\in\Ksc$, the 
$$\by^T\log\{\cosh(\smhalf\bc_{\qDens(\balpha|\kappa)})\}\quad\mbox{sufficient statistic is incremented by}\quad \ynew\log\{\cosh(\smhalf\calphakappanew)\}$$
where
$$\calphakappanew=\sqrt{\bcnew^T
\bSigma_{\qDens(\bbeta,\bu|\kappa)}\bcnew+
\{\bcnew^T\bmu_{\qDens(\bbeta,\bu|\kappa)}-\log(\kappa)\}^2}.$$
and $\bmu_{\qDens(\bbeta,\bu|\kappa)}$ and 
$\bSigma_{\qDens(\bbeta,\bu|\kappa)}$ are the current
$\qDens$-density parameters of $(\bbeta,\bu)|\kappa$.
Continuing in this fashion, we arrive at 
Algorithm \ref{alg:ONLalgo} for real-time count
response semiparametric regression, and only
requiring low-dimensional sufficient statistics 
storage and updating.

\setcounter{algorithm}{1}
%
\begin{algorithm}[!th]
\begin{center}
\begin{minipage}[t]{160mm}
\hrule
\begin{small}
\begin{itemize}
\item[]\null 
\begin{itemize}
\item[] \null
\begin{itemize}
\item[] $\logML(\kappa)\longleftarrow
\smhalf\bmu_{\qDens(\bbeta,\bu|\kappa)}^T\big(\CTy-\kappa\,\CTone\big)
-\by^T\log\left\{
\cosh\Big(\smhalf c_{\qDens(\balpha|\kappa)}\Big)\right\}$
\item[] $\qquad\qquad-\kappa\bone^T\log\left\{
\cosh\Big(\smhalf c_{\qDens(\balpha|\kappa)}\Big)\right\}
-\displaystyle{
\frac{\Vert\bmu_{\qDens(\bbeta|\kappa)}\Vert^2
+\tr(\bSigma_{\qDens(\bbeta|\kappa)})}{2\sigsqbeta}}
+\smhalf\log|\bSigma_{\qDens(\bbeta,\bu|\kappa)}|$ 
\item[]$\qquad\qquad+\bone^T\log\{\Gamma(\by+\kappa\bone)\}
+n\big[\smhalf\kappa\log(\kappa)-\log(2)\kappa-\log\{\Gamma(\kappa)\}\big]
-\smhalf\log(\kappa)(\yTone)$
\item[] For $j = 1,\ldots,r:$
\begin{itemize}
\item[] $\logML(\kappa)\longleftarrow \logML(\kappa)
+\mu_{\qDens(1/\sigma_j^2|\kappa)}
\big\{\lambda_{\qDens(\sigma^2_j|\kappa)}
-\mu_{\qDens(1/a_j|\kappa)}
-\smhalf\Vert\bmu_{\qDens(\bu_j|\kappa)}\Vert^2
-\smhalf\tr\big(\bSigma_{\qDens(\bu_j|\kappa)}\big)\big\}$
\item[] $\qquad\qquad+\mu_{\qDens(1/a_j|\kappa)}
\big(\lambda_{\qDens(a_j|\kappa)}-\ssigma^{-2}\big)
-\smhalf(K_j+1)\log\big(\lambda_{\qDens(\sigma^2_j|\kappa)}\big)
-\log\big(\lambda_{\qDens(a_j|\kappa)}\big)$
\end{itemize}
\item[] $\qDens^*(\kappa)\longleftarrow\pDens(\kappa)\exp\{\logML(\kappa)\}\Big/
{\displaystyle\sum_{\kappa'\in\Ksc_n}}\pDens(\kappa')\exp\{\logML(\kappa')\}$
\end{itemize}

\item[] $\qDens^*(\bbeta,\bu)\longleftarrow
{\displaystyle\sum_{\kappa\in\Ksc_n}}\Bigg[\qDens^*(\kappa)
|2\pi\bSigma_{\qDens(\bbeta,\bu|\kappa)}|^{-1/2}$\\
$\null\qquad\qquad\qquad\times
\exp\left\{-\smhalf\left(
\bbetabuVec-\bmu_{\qDens(\bbeta,\bu|\kappa)}\right)^T 
\bSigma_{\qDens(\bbeta,\bu|\kappa)}^{-1}    
\left(\bbetabuVec-\bmu_{\qDens(\bbeta,\bu|\kappa)}\right)
\right\}\Bigg]
$
\item[] For $j = 1,\ldots,r:$
\begin{itemize}
\item[] $\qDens^*(\sigma_j^2)\longleftarrow 
{\displaystyle\sum_{\kappa\in\Ksc_n}
\left\{
\frac{\qDens^*(\kappa)\lambda_{\qDens(\sigma^2_j|\kappa)}^{(K_j+1)/2}}
{\Gamma\big(\smhalf(K_j+1)\big)}\right\}
(\sigma_j^2)^{-(K_j+1)/2-1}
\exp\left(-\frac{\lambda_{\qDens(\sigma^2_j|\kappa)}}{\sigma_j^2}\right)}.
\ \ \sigma_j^2>0,
$
\end{itemize}
\item[] Produce summaries based on $\qDens^*(\bbeta,\bu)$, $\qDens^*(\kappa)$
and $\qDens^*(\sigma_j^2)$,\ $1\le j\le r$.
\end{itemize}
\item[] until data no longer available or analysis terminated.
\end{itemize}
\end{small}
\hrule
\end{minipage}
\end{center}
\caption{\textbf{continued.}\ \textit{This is a continuation of the description of this algorithm that
commences on a preceding page.}}
\end{algorithm}
%

\subsection{The Step of Algorithm \ref{alg:ONLalgo} 
Involving the Reduced Atom Set $\Ksc_n$}\label{sec:atomsRedu}

When developing and testing Algorithm \ref{alg:ONLalgo} we first 
looked into using the original atom set $\Ksc$ during the online updates.
However, when $\Ksc$ is kept fixed, there is a tendency for the $\kappa$ probability 
mass to pile towards the left or right extremities of $\Ksc$ which, in turn, 
adversely impacts the quality of the $\qDens^*(\bbeta,\bu)$ and 
$\qDens^*(\sigma^2_j)$ approximations. We do not have an explanation
for the occurrence of this phenomenon. Some experimentation showed that
the online $\qDens$-densities of the main model parameters could 
achieve similar behaviour to their batch counterparts if the 
$\kappa$ atoms were sequentially reduced and concentrated around the
posterior mean of $\qwarm^*(\kappa)$. Further research on this 
aspect seems warranted. However, with practicality in mind, we devised
the following simple scheme for achieving such concentration.
Given the skewed nature of posterior distributions of positive-valued
parameters, we work with logarithm of $\kappa$.
Let $\lambda\equiv\log(\kappa)$ and let $\qwarm^*(\lambda)$ be the 
probability mass function of $\lambda$ corresponding to $\qwarm^*(\kappa)$.
Next let $\muwarm^{\lambda}$ and $\sigmawarm^{\lambda}$ denote the 
mean and standard deviation of $\qwarm^*(\lambda)$. Asymptotic normality
considerations dictate that most of the probability mass of $\qwarm^*(\lambda)$
is in the interval 
$$\big(\muwarm^{\lambda}-\tau\sigmawarm^{\lambda}, 
\muwarm^{\lambda}+\tau\sigmawarm^{\lambda}\big)
\quad\mbox{where}\quad\tau\approx 3.
$$
For $n\ge\nwarm$ a reasonable away to coax $\qDens^*(\kappa)$ to be more 
concentrated around $\qwarm^*(\kappa)$ is to set 
\begin{equation}
\Ksc_n=\left\{\kappa\in\Ksc: 
\muwarm^{\lambda}-\tau\sigmawarm^{\lambda}\sqrt{\nwarm/n}
\le\log(\kappa)\le 
\muwarm^{\lambda}+\tau\sigmawarm^{\lambda}\sqrt{\nwarm/n}
\right\},
\label{eq:KscnFormula}
\end{equation}
which is also based on standard asymptotic normality considerations.
In our simulated data assessment of Algorithm \ref{alg:ONLalgo}, 
we adopted (\ref{eq:KscnFormula}) with $\tau=3.5$. Lastly, for very large $n$,
we need to guard against $\Ksc_n$ becoming null. We suggest to 
cease the atoms reduction when the number of atoms reaches a low
value such as $5$.


\section{Numerical Results}\label{sec:numRes}

We now evaluate and illustrate the performance of Algorithms 
\ref{alg:BCHalgo} and \ref{alg:ONLalgo} using both simulated
and actual data. Section \ref{sec:simuAlgOne} uses the
same simulated data setting as Luts \myand Wand (2015)
to assess comparative accuracy and speed of 
Algorithm \ref{alg:BCHalgo} against a Markov chain
Monte Carlo benchmark. In Section \ref{sec:simuAlgTwo}
we conduct a simulation-based assessment of 
Algorithm \ref{alg:ONLalgo}. In this study the
real-time fits are compared with the more computationally
expensive batch fits. As with any simulation study,
the results of Sections \ref{sec:simuAlgOne}
and Section \ref{sec:simuAlgTwo} are necessarily
limited in that they can only treat a few specific
scenarios. In Section \ref{sec:pollenApplic}
we illustrate use of the methodology
for some actual data concerning pollen counts.

\subsection{Simulated Data Assessment of Algorithm \ref{alg:BCHalgo}}\label{sec:simuAlgOne}

Our simulated data assessment of Algorithm \ref{alg:BCHalgo} involved
data generated according to the following Negative Binomial additive
model:
$$y_i|x_{i1}, x_{i2} \simind \mbox{Negative-Binomial}
\big(\exp\{\etaTrueOne(x_{1i}) + \etaTrueTwo(x_{2i})\}\big),\kappaTrue),
\quad 1\le i\le 500$$
where
$$
\etaTrueOne(x)=\cos(4\pi x) + 2x,\quad
\etaTrueTwo(x)=0.4 \phi(x; 0.38,0.08) - 1.02x 
+ 0.018x^2 + 0.08 \phi(x;0.75,0.03)
$$
and $\phi(x;\mu,\sigma)$ denotes the density function of the Normal 
distribution with mean $\mu$ and standard deviation $\sigma$, evaluated at $x$. 
We set $\kappaTrue=3.8$, which corresponds to a relatively high amount of
over-dispersion. The predictor data were generated according to
$$x_{i1},x_{i2}\simind\mbox{Uniform}(0,1),\quad 1\le i\le 500.$$
We used the following penalized spline model for
estimation of $\etaTrueOne(x_1) + \etaTrueTwo(x_2)$:
\begin{equation}
\beta_0 + \beta_1x_1 + \beta_2 x_2 + \sum_{k=1}^{K_1} 
u_{1k} z_{1k}(x_1) + \sum_{k=1}^{K_2} u_{2k} z_{2k}(x_2),
\quad u_{jk}\simind N(0,\sigma_j^2),\quad j=1,2.
\label{eq:laughIn}
\end{equation}
Here $z_{1k}$ and $z_{2k}$ are canonical O'Sullivan spline bases
as described in Section 4 of Wand \myand Ormerod (2008).
The basis sizes were $K_1=K_2=17$. This set-up is an $r=2$ 
special case of model (\ref{eq:negBinModel}) with, for example,
\begin{equation}
\bX=\big[1\ x_{1i}\ x_{2i}\big]_{1\le i\le 500}
\quad\mbox{and}\quad
\bZ=\big[z_{11}(x_{1i})\cdots z_{1K_1}(x_{1i})\ 
z_{21}(x_{2i})\cdots z_{2K_2}(x_{2i})\big]_{1\le i\le 500}.
\label{eq:Zcars}
\end{equation}
We imposed the prior distributions:  
$$\beta_0,\beta_1,\beta_2\simind N(0,10^5),\quad\sigma_1,\sigma_2\simind\mbox{Half-Cauchy}(10^5)
\quad\mbox{and}\quad
\pDens(\kappa)\propto\exp(-\kappa/100),\quad\kappa\in\Ksc,$$
where the atom set $\Ksc$ is a geometric sequence of size 50 
between $\kappa^{\text{true}}/10$ and  $10\kappa^{\text{true}}$. 
We fitted the special case of model (\ref{eq:negBinModel}),
corresponding to (\ref{eq:laughIn}) and (\ref{eq:Zcars}),
via Algorithm \ref{alg:BCHalgo}. The convergence was 
assessed by monitoring the relative change in 
$\log\{\underline{p}(\by; q|\kappa)\}$, with a stopping criterion set 
at $10^{-10}$. One hundred simulation replications were
performed. 

To facilitate accuracy assessment we also obtained fits based
on a Markov chain Monte Carlo approach.
In order to allow a fair comparison 
between the two approaches, $\kappa$ was restricted
to the same finite set $\Ksc$.
Markov chain Monte Carlo fitting and inference was achieved
using the \textsf{JAGS} Bayesian inference engine via the 
\textsf{R} package \textsf{rjags} (Plummer, 2025).
For each replication, chains of length $10000$ were obtained. 
The first $5000$ values were discarded as burn-in 
Then, thinning by a factor of $5$ was applied, 
leading to a retained samples of size of $1000$. 

\subsubsection{Assessment of Accuracy}

The accuracy score for a structured mean field variational Bayes 
approximate posterior density function $\qDens^*(\theta)$
of a generic continuous parameter $\theta$ is defined as
\begin{equation}
\text{accuracy}(\qDens^*) = 100\left(1-\frac{1}{2}\int_{-\infty}^{\infty} 
\big|\qDens^*(\theta) - \pDens(\theta|\by)\big|d\theta\right)\%.
\label{eq:accDefn}
\end{equation}
For the discrete parameter $\kappa$, an analogous definition applies
with the integral replaced by the sum over $\Ksc$. 
The density $\pDens(\theta|\by)$ was estimated from the 
Markov chain Monte Carlo samples for $\theta$ using the \texttt{bkde()} and 
\texttt{dpik()} kernel density estimation 
functions within the \textsf{R} package \textsf{KernSmooth}
(Wand \myand Ripley, 2024).

%
\begin{figure}
\centering
\includegraphics[width=\textwidth]{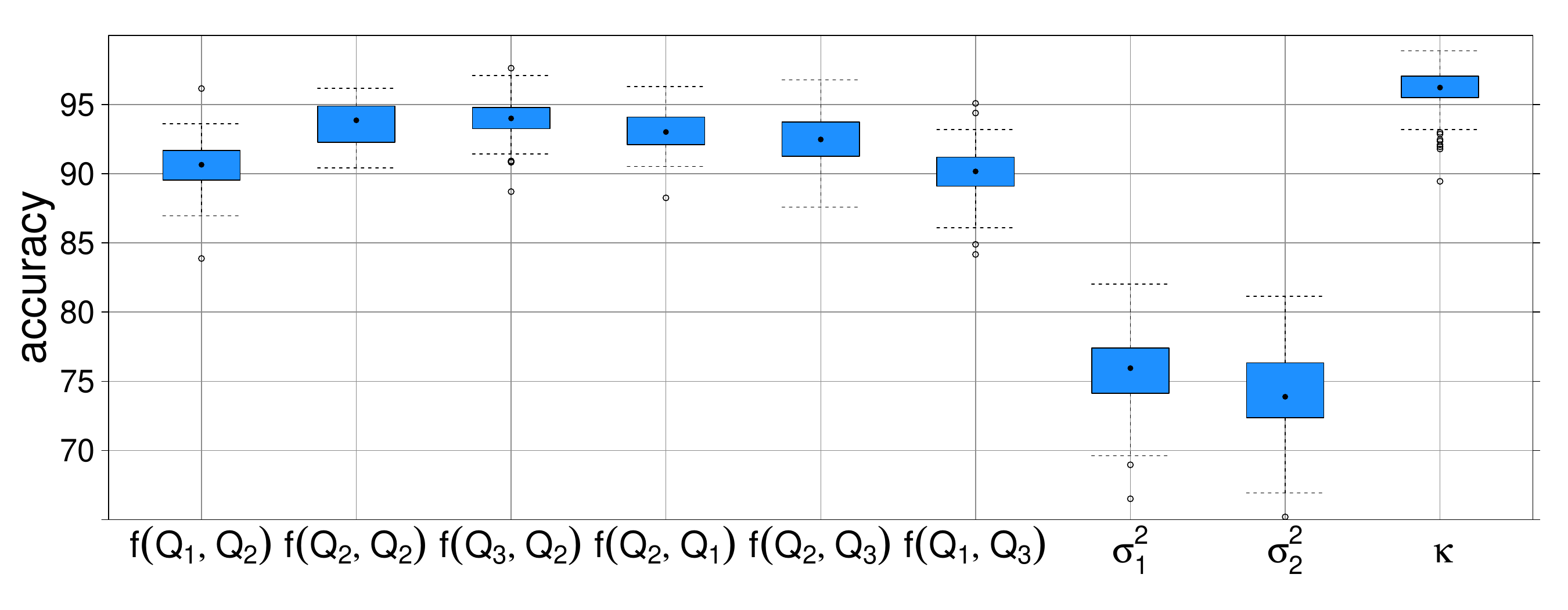}
\caption{\textit{Boxplots of accuracy scores, as defined by (\ref{eq:accDefn}), 
for the Algorithm \ref{alg:BCHalgo} simulation study.
The $f(Q_k,Q_{k'})$ notation is defined by (\ref{eq:fdefn}) and 
subsequent text.}}
\label{fig:accBoxplots}
\end{figure}

Figure \ref{fig:accBoxplots} displays the boxplots of the accuracy scores 
for estimation of the function 
\begin{equation}
f(x_1,x_2)\equiv\exp\{\etaTrueOne(x_1)+\etaTrueTwo(x_2)\}
\label{eq:fdefn}
\end{equation}
evaluated at the sample quartiles of the $x_{1i}$ and $x_{2i}$ values.
We use the notation $Q_k$, $k=1,2,3$, to denote the quartiles.
Accuracy scores for $\sigma^2_1, \sigma^2_2$ and $\kappa$ were also obtained. 
The boxplots in Figure \ref{fig:accBoxplots} indicate satisfactory accuracies
for this simulation setting. When compared with Figure 4 of Luts \myand Wand (2015),
improvements over that article's semiparametric mean field variational Bayes
approach are observed, with the most notable gain being for $\kappa$.

%
\begin{figure}
\centering
\includegraphics[width=\textwidth]{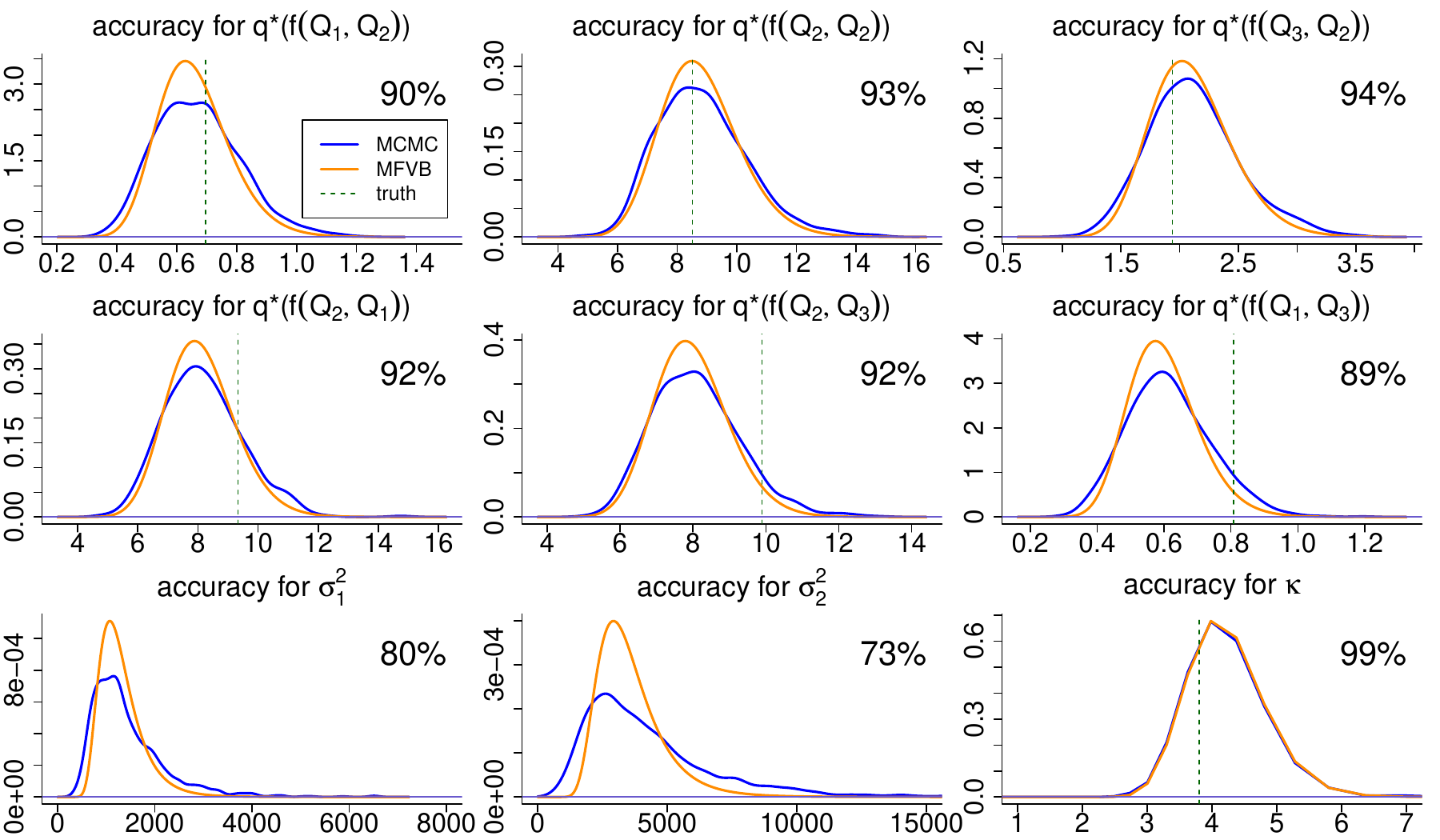}
\caption{
\textit{Illustrations of the accuracy of the structured mean field variational
Bayes (MFVB) posterior density functions and probability mass function approximations
obtained from Algorithm \ref{alg:BCHalgo}. In each panel, the 
MFVB approximate density functions or probability mass function 
for a quantity of interest is compared with its
Markov chain Monte Carlo (MCMC) counterpart.
The percentage is the accuracy score according to (\ref{eq:accDefn}).
The vertical lines indicate true values according to the simulation set-up.
}}
\label{fig:postDensCompar}
\end{figure}

Figure \ref{fig:postDensCompar} shows the accuracy costs of the mean field approximation by
comparing the $\qDens$-density functions with those based on the Markov chain
Monte Carlo for a randomly chosen replication. 
To aid visualisation we replaced the $\kappa$ probability
mass functions by polygons formed by joining each of the the atom/probability
pairs. The polygons were then normalized to have areas under the curve
equal to $1$. It can be seen that variational approximate posterior densities 
are nearly as wide as the Markov chain Monte Carlo counterparts for 
the functions evaluated at the quartiles, and also centered about true values.
However, some inaccuracy is apparent. The discrete densities of $\kappa$ obtained via
Algorithm \ref{alg:BCHalgo} and Markov chain Monte Carlo 
and are very similar to each other and centered around $\kappa^{\text{true}}$.

\subsubsection{Assessment of Speed}

The execution time of Markov chain Monte Carlo represents a significant 
bottleneck in Bayesian analysis, which is one of the main motivations for 
exploring faster alternatives such as variational approximations. 
Algorithm \ref{alg:BCHalgo} requires achieving convergence of 
$|\Ksc|$ variational algorithms, where $|\Ksc|$ is the cardinality of $\Ksc$.
We recommend the use of warm starts when marching, in order, through the $\kappa$ atoms
to accelerate convergence and thus reduce processing time. 
The overall computational burden is linear in $|\Ksc|$, 
and appropriate choice of the $\kappa$ prior can help prevent excessively long run times.
That said, our method achieves very good speed. The average (standard deviation) elapsed time of the 
two methods across all 100 simulated datasets is $117.8$ $(1.876)$ seconds
for the Markov chain Monte Carlo approach, and $2.088$ 
$(0.1440)$ seconds for Algorithm \ref{alg:BCHalgo}. The simulations were 
run in \textsf{R} (\textsf{R} Core Team, 2025), version 4.4.3, on a machine 
with $12$ cores and $24$ gigabytes of random access memory.

\subsection{Simulated Data Assessment of Algorithm \ref{alg:ONLalgo}}\label{sec:simuAlgTwo}

To assess the efficacy of Algorithm \ref{alg:ONLalgo} we simulated data
sets corresponding to the Negative binomial nonparametric regression model
\begin{equation}
y_i|x_i\simind \mbox{Negative-Binomial}\big(\etaTrue(x_i),\kappaTrue\big),
\quad 1\le i\le\nFull,
\label{eq:coldOolong}
\end{equation}
where
$$\etaTrue(x)\equiv 0.3\phi(x;0.2,0.08) - 0.3\phi(x;0.65,0.23) + 0.4\phi(x;0.45,0.08)$$
where $\phi(\cdot;\mu,\sigma)$ is as defined earlier in this section. 
In (\ref{eq:coldOolong}) $\nFull$ signifies the full real-time data
sample size. In all of our examples we set $\nFull=1000$.
The predictor data were generated according to $x_i\simind\mbox{Uniform}(0,1)$.

Bayesian penalized splines of the form
$$\beta_0+\beta_1\,x+\sum_{k=1}^K u_k z_k(x),\quad u_k|\sigma^2\simind N(0,\sigma^2),$$
were used to model and estimate $\etaTrue$. The $z_k$ spline basis functions
are analogous to those described in Section \ref{sec:simuAlgOne}. 
In this real-time example we used 35 interior knots, which entails 
use of $K=37$ basis functions. 
This set-up corresponds to the $r=1$ special
case of Algorithms \ref{alg:BCHalgo} and \ref{alg:ONLalgo}.
We also imposed the prior distributions: 
$$\beta_0,\beta_1\simind N(0,10^5),\quad\sigma\sim\mbox{Half-Cauchy}(10^5)
\quad\mbox{and}\quad
\pDens(\kappa)\propto\exp(-\kappa/100),\quad\kappa\in\Ksc$$
where $\Ksc$ is the geometric sequence of length 50 between
$\kappaTrue/10$ and $10\kappaTrue$.

Real-time data scenarios based on (\ref{eq:coldOolong}) were simulated 
for each of 
$$\kappaTrue\in\{5,10,20,40\},$$
corresponding to count responses with varying amounts of overdispersion.
Three replications of $(x_i,y_i)$, $1\le i\le\nFull$, data were generated 
within the \textsf{R} computing environment according to the command
\texttt{set.seed(s)} with \texttt{s} set to each of $1$, $2$ and $3$.
These data were then fed into Algorithm \ref{alg:ONLalgo} with samples of
size $n=\nwarm,\nwarm+1,\ldots,\nFull$. In all but one case we used
$\nwarm=100$ and achieved good convergence.
The exception was $\kappaTrue=20$ with the third seed value, in which
case the longer warm-up of $\nwarm=200$ was warranted.
The same sequential data sets were fed into Algorithm \ref{alg:BCHalgo} to allow comparison between
the online fits and those obtained via ordinary batch processing.

\begin{figure}[!h]
\centering
\includegraphics[width=0.95\textwidth]{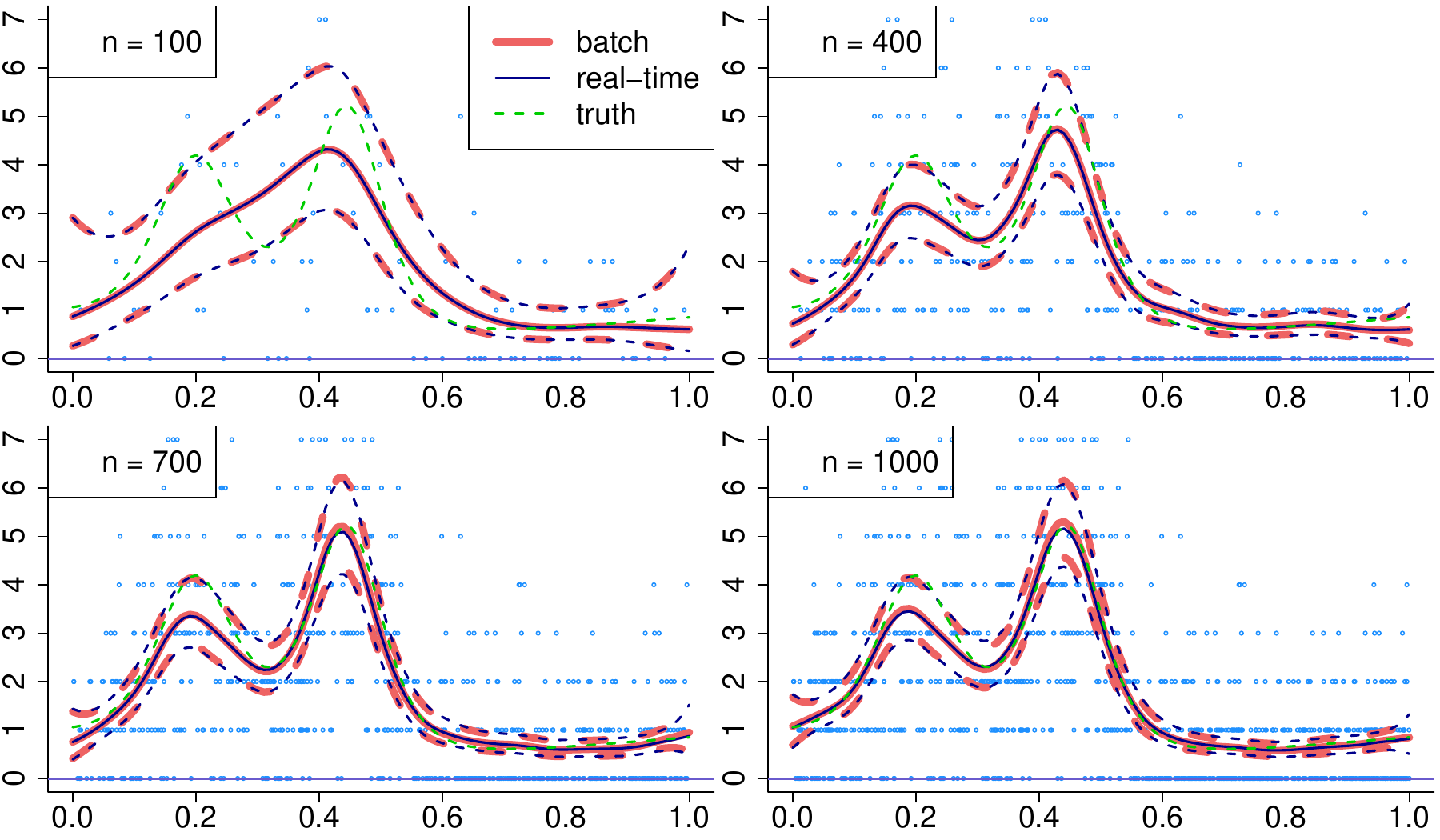}
\caption{\textit{Some illustrative comparisons for $\kappaTrue=5$
between the real-time Negative Binomial nonparametric regression estimates 
based on Algorithm \ref{alg:ONLalgo} with the batch counterparts based on 
Algorithm \ref{alg:BCHalgo}. The solid curves correspond to posterior means. 
The dashed curves correspond to pointwise approximate 95\% credible intervals.
The scatterplots correspond to the current regression data.}}
\label{fig:NegBinNPregStills}
\end{figure}

The results of our simulation-based assessment of Algorithm \ref{alg:ONLalgo}
are presented as movies within the supplementary material.
\footnote{For this pre-publication version of this article, the movies
are on the following web-site:\\
$\qquad\qquad$\texttt{https://matt-p-wand.net/M+Wmovies.html}}
Figure \ref{fig:NegBinNPregStills} shows some of the 
frames from the first $\kappaTrue=5$ movie.

Figure \ref{fig:NegBinNPregStills} and, in particular, the movies
show that Algorithm \ref{alg:ONLalgo}'s real-time inference for the 
mean structure is quite similar to that obtained with successive
batch fitting. Clearly, the former is preferable from a speed
standpoint in streaming data applications. High quality real-time
inference for the $\kappa$ nuisance parameter appears to be out of reach. 
The strategy used in Figure \ref{fig:NegBinNPregStills} 
and the movies, and described in Section \ref{sec:atomsRedu}, 
aims to ensure that the running approximate posterior distributions of $\kappa$ 
are reasonable enough to not adversely impact real-time estimation
of the mean structure.

\subsection{Application to Pollen Counts Data}\label{sec:pollenApplic}

This application involves data daily ragweed pollen
counts in Kalamazoo, U.S.A., during the 1991--1994 ragweed seasons.
The data correspond to the study described in Stark \textit{et al.} (1997). 
The model is of the form 
\begin{equation}
y_i\simind\mbox{Negative-Binomial}
\big\{\exp\big(\beta_0+\beta_1\,x_{1i}+\beta_2\,x_{2i}+\beta_3\,x_{3i}
+\eta_{z_i}(x_{4i})\big),
\kappa\big\},
\quad 1\le i\le n,
\label{eq:ragPoissMod}
\end{equation}
where $n=334$ corresponds to the total number of days when ragweed pollen
was in season during 1991--1994. The variables in (\ref{eq:ragPoissMod}) are
ragweed pollen count on the $i$th day ($y_i$),
temperature residual on the $i$th day ($x_{1i}$),
indicator of significant rain on the $i$th day ($x_{2i}$),      
wind speed in knots on the $i$th day ($x_{3i}$),
day number of ragweed pollen season for the current year on which $y_i$ was recorded ($x_{4i}$)
and a categorical variable for the year in which $y_i$ was recorded 
(one of 1991, 1992, 1993 or 1994) ($z_i$).
Here temperature residuals are the residuals from fitting penalized
splines, each having 5 effective degrees of freedom, to temperature
(in degrees Fahrenheit) versus day number for each 
ragweed pollen season. Mixed model-based penalized splines
were used for modelling the $\eta_{z}$, $z\in\{1991,1992,1993,1994\}$. 
The variance parameters used in each of the four penalized spline
components are $\sigma^2_1$, $\sigma_2^2$, $\sigma_3^2$ and $\sigma_4^2$. 
The full model is
\begin{equation}
\begin{array}{c}
y_i|\bbeta,\bu,\kappa\simind 
\mbox{Negative-Binomial}\big(\exp\{(\bX\bbeta+\bZ\bu)_i\},\kappa\big),\quad
\bbeta\sim N(\bzero,10^{10}\bI)
\\[1ex]
\bu|\sigma_1^2,\ldots,\sigma_4^2\sim N\big(\bzero,\mbox{blockdiag}
(\sigma_1^2\,\bI_{K_1},\ldots,\sigma_4^2\,\bI_{K_4})\big),\\[1ex]
\sigma_1,\ldots,\sigma_4\simind
\mbox{Half-Cauchy}(10^5),\quad
\pDens(\kappa)\propto \exp(-\kappa/100),\quad\kappa\in\Ksc
\end{array}
\label{eq:ragweedModel}
\end{equation}
where $\Ksc$ is a geometric sequence of length 100 between $0.5$ to $50$.
The design matrices in (\ref{eq:ragweedModel}) are
$$
\bX=\left[
{\setlength\arraycolsep{1pt}
\begin{array}{ccccccccc}
1&\ x_{11}&\cdots&x_{41}&\ \ I(z_1\!\!=\!\!1992)&\ \ x_{41}I(z_1\!\!=\!\!1992)&\ \cdots\ &I(z_1\!\!=\!\!1994)&\ \ x_{41}I(z_1\!\!=\!\!1994)\\
\vdots&\vdots&\vdots&\vdots&\vdots&\vdots&\vdots&\vdots&\vdots\\
1&\ x_{1n}&\cdots&x_{4n}&\ \ I(z_n\!\!=\!\!1992)&\ \ x_{4n}I(z_n\!\!=\!\!1992)&\ \cdots\ &I(z_n\!\!=\!\!1994)&\ \ x_{4n}I(z_n\!\!=\!\!1994)\\
\end{array}
}
\right]
$$
and $\bZ=[\bZ_{1991}\,\bZ_{1992}\,\bZ_{1993}\,\bZ_{1994}]$ where $\bZ_{1991}$ is an $n\times K_j$ matrix
with $(i,k)$ entry equal to $I(z_i=1991)z_k(x_{4i})$ and $\bZ_{1992},\ldots,\bZ_{1994}$ are defined
analogously. The $z_k$ basis functions are of the same type used earlier in
this section.
The $\bbeta$ and $\bu$ vectors contain the coefficients to match the columns of $\bX$ and $\bZ$
respectively. Lastly, the spline basis sizes were  $K_1=K_2=K_3=17$ and $K_4=16$.
This difference in the spline basis sizes is due to the numbers
days in the ragweed pollen season varying in length 
between the four years.  They range from $78$ for year 1994 to $92$ to year 1991.

%
\begin{figure}[b]
\centering
\includegraphics[width=1.0\textwidth]{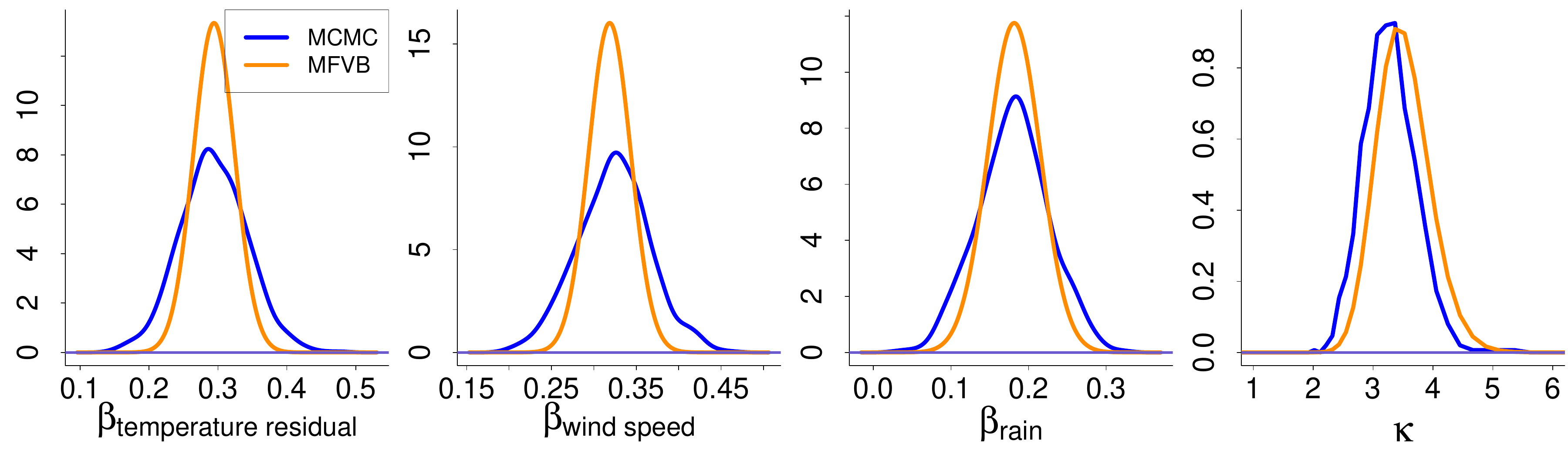}
\caption{\textit{Comparison of posterior density function
and probability mass function approximations
based on structured mean field variational Bayes (MFVB) and 
Markov chain Monte Carlo (MCMC) for four of the
parameters in model (\ref{eq:ragweedModel}).}}
\label{fig:densRagweed}
\end{figure}

Model (\ref{eq:ragweedModel}) is an $r=4$ special case of (\ref{eq:negBinModel}).
We used Algorithm \ref{alg:BCHalgo} to perform approximate
Bayesian inference.

\begin{figure}[!]
\centering
\includegraphics[width=1.0\textwidth]{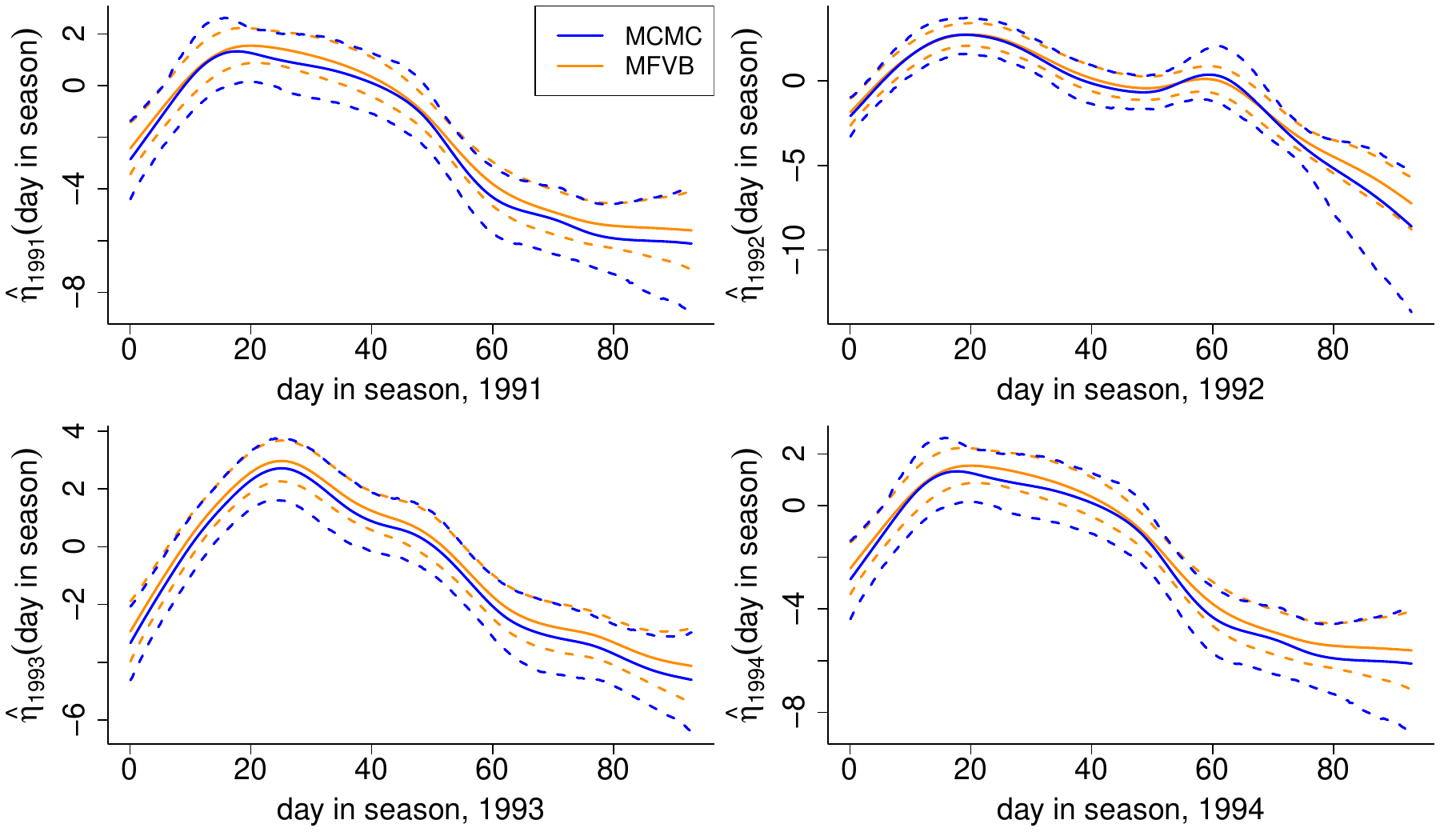}
\caption{\textit{Structured mean field variational Bayes (MFVB) and Markov chain Monte
Carlo (MCMC) posterior means (solid lines) and 95\% pointwise credible intervals 
(dashed lines) for estimation of the functions $\eta_{1991},\ldots, \eta_{1994}$ 
in the model conveyed by (\ref{eq:ragPoissMod}) and (\ref{eq:ragweedModel}).}}
\label{fig:meanCurvesRagweed}
\end{figure}

%
Figure \ref{fig:densRagweed} shows the posterior density functions for the parameters 
associated with the linear effects of the quantitative explanatory variables considered 
in the model. On the right, the posterior probability distribution function of $\kappa$ 
is represented. It can be noticed that all the the coefficient parameters are 
significantly different from zero.
The variational approximate density functions for the linear effects 
appear narrower than the Markov chain Monte Carlo ones; 
the approximate posterior distribution of $\kappa$ appears slightly shifted to the right 
compared to the exact one. Both distributions assign posterior mass to values ranging 
from $2$ to $5$, supporting the suitability of the Negative-Binomial response model. 

The trend of the linear predictor over the days of the season, shown separately for 
each year, is displayed in Figure \ref{fig:meanCurvesRagweed}. Solid lines indicate the posterior 
mean, and dashed lines the posterior pointwise credible intervals at level 95\%. The 
four curves exhibit similar behavior, with a peak around the 20th day of the season, 
followed by a decreasing trend.

\section{Conclusion}\label{sec:conclusion}

Our convex solution to the count response semiparametric regression problem
described here offers stability and real-time processing advantages. 
The inferential accuracy is often reasonable. However, it also prone
to some inaccuracy and this aspect needs to be taken into account
when trading off against speed. Even though we have focussed on 
semiparametric regression models, the same general approach 
applies to numerous other count response settings.

\section*{Acknowledgement}

We are grateful to Emanuele Aliverti for discussions related to this
article. This research was partially supported by Australian Research 
Council grant DP230101179. 

\section*{References}

\bib
Bishop, C.M. (2006). {\it Pattern Recognition and Machine Learning.}
New York: Springer.

\bib
Durante, D. \myand Rigon, T. (2019).
Conditionally conjugate mean-field variational Bayes
for logistic models. \textit{Statistical Science},
\textbf{34}, 472--485.

\bib
Jaakkola, T.S. \myand Jordan, M.I. (2000). Bayesian
parameter estimation via variational methods.
{\it Statistics and Computing}, {\bf 10}, 25--37.

\bib
Lee, C.Y.Y. and Wand, M.P. (2016).
Variational inference for fitting complex Bayesian mixed 
effects models to health data. \textit{Statistics in Medicine},
{\bf 35}, 165--188.

\bib
Luts, J., Broderick, T. and Wand, M.P. (2014).
Real-time semiparametric regression.
\textit{Journal of Computational and Graphical Statistics},
{\bf 23}, 589--615.

\bib
Luts, J. and Wand, M.P. (2015).
Variational inference for count response semiparametric regression.
\textit{Bayesian Analysis}, {\bf 10}, 991--1023.

\bib
Miao, Y., Kook, J.H., Lu, Y., Guindani, M. \myand Vannucci, M. (2020).
Scalable Bayesian variable selection models for count data.
In Y. Fan, D.J. Nott, M.S. Smith, J.-C. Dortet-Bernadet,  editors,
\textit{Flexible Bayesian Modelling},
pp. 187--219, London: Academic Press.

\bib
Ormerod, J.T. and Wand, M.P. (2010).
Explaining variational approximations.
{\it The American Statistician},
{\bf 64}, 140--153.

\bib
Pillow, J.W. \myand Scott, J.G. (2012).
Fully Bayesian inference for neural models with 
negative-binomial spiking. 
In Bartlett, P., Pereira, F.C.N., Burges, C.J.C.,
Bottou, L. \myand Weinberger, K.Q.,
editors,
In \textit{Advances in Neural Information Processing Systems 25},
pp.\ 1898--1906.
Cambridge, Massachusetts: MIT Press.

\bib
Plummer, M. (2025). \textsf{rjags}: Bayesian graphical models using 
Markov chain Monte Carlo. \textsf{R} package version 4-17.
\texttt{CRAN.R-project.org/package=rjags}.

\bib
Polson, N.G., Scott, J.G. \myand Windle, J. (2013).
Bayesian inference for logistic models using P\'olya-Gamma
latent variables. \textit{Journal of the American Statistical
Association}, \textbf{108}, 1339--1349.

\bib
\textsf{R} Core Team (2025). \textsf{R}: A language
and environment for statistical computing.
\textsf{R} Foundation for Statistical Computing, Vienna, Austria.
\texttt{https://www.R-project.org/}.

\bib
Saul, L.K. \myand Jordan, M.I. (1996).
Exploiting tractable substructures in intractable
networks. In Mozer, M.C., Jordan, M.I. \myand Petsche T.,
editors, \textit{Advances in Neural Information Processing
Systems 9}, pp. 435--442. Cambridge, Massachusetts: MIT Press.

\bib
Stark, P.C., Ryan, L.M., McDonald, J.L. \myand Burge, H.A. (1997).
Using meteorologic data to model and predict daily ragweed
pollen levels. \textit{Aerobiologia}, \textbf{13}, 177--184.

\bib
Wainwright, M.J. \myand Jordan, M.I. (2008).
Graphical models, exponential families and variational 
inference. \textit{Foundations and Trends in 
Machine Learning}, {\bf 1}, 1--305.

\bib
Wand, M.P. and Ormerod, J.T. (2008).
On semiparametric regression with O'Sullivan penalized splines.
{\it Australian and New Zealand Journal of Statistics},
{\bf 50}, 179--198.

\bib
Wand, M.P., Ormerod, J.T., Padoan, S.A. \myand
Fr\"uhwirth, R. (2011).
Mean field variational Bayes for elaborate distributions.
{\it Bayesian Analysis}, {\bf 6}, 847--900.

\bib
Wand, M. \myand Ripley, B. (2024). \textsf{KernSmooth}: Functions for
kernel smoothing supporting Wand \myand Jones (1995). 
\textsf{R} package version 4-17. \texttt{CRAN.R-project.org/package=KernSmooth}.

\bib
Wand, M.P. and Yu, J.C.F. (2022).
Density estimation via Bayesian inference engines.
\textit{Advances in Statistical Analysis}, \textbf{106}, 199--216.

\bib
Zhao, Y., Staudenmayer, J., Coull, B.A. and Wand, M.P. (2006).
General design Bayesian generalized linear mixed models.
{\it Statistical Science}, {\bf 21}, 35--51.

\bib
Zhou, M. Li, L., Dunson, D. \myand Carin, L. (2012).
Lognormal and Gamma mixed Negative Binomial regression.
In \textit{
Proceedings of the 29th International Conference on Machine Learning},
pp.\ 1343--1350.

\null
\vfill\eject
%
%
\renewcommand{\theequation}{S.\arabic{equation}}
\renewcommand{\thesection}{S.\arabic{section}}
\renewcommand{\thetable}{S.\arabic{table}}
\setcounter{equation}{0}
\setcounter{table}{0}
\setcounter{section}{0}
\setcounter{page}{1}
\setcounter{footnote}{0}

\centerline{\Large Supplement for:}
\vskip5mm
\centerline{\Large\bf Variational Inference for Count Response}
\vskip2mm
\centerline{\Large\bf Semiparametric Regression: A Convex Solution}
\vskip5mm
\centerline{\normalsize\sc Virginia Murru and Matt P. Wand}
\vskip5mm
\centerline{\textit{\myUniversita\ di Padova and University of Technology Sydney}}
\vskip5mm{\null}

\section{P\'olya-Gamma Distribution Definitions and Results}\label{sec:PGdist}

The P\'olya-Gamma distribution plays an important role
in this article's variational inference approach. 
In this section we provide relevant definitions and results.

\subsection{The Gamma Distribution}

A random variable $x$ has a Gamma distribution with shape parameter
$\alpha>0$ and rate parameter $\beta>0$, written
$$x\sim\mbox{Gamma}(\alpha,\beta),$$
if and only if its probability density function is 
$$\pDens(x;\alpha,\beta)=\frac{\beta^{\alpha}}{\Gamma(\alpha)}\,x^{\alpha-1}
\exp(-\beta x),\quad x>0.$$

\subsection{The P\'olya-Gamma Distribution}\label{sec:PGdistn}

A random variable $x$ has a P\'olya-Gamma distribution with shape
parameter $b>0$ and tilting parameter $c>0$, written
$$x\sim\mbox{P\'olya-Gamma}(b,c),$$
if and only if 
$$x\quad\mbox{is equal in distribution to}\quad\frac{1}{2\pi^2}\sum_{k=1}^{\infty}
\frac{g_k}{(k-1/2)^2 + c^2/(4\pi^2)}
$$
where
$$g_k\simind\mbox{Gamma}(b,1),\quad k=1,2,\ldots$$
Note that there is no closed form expression for the density function
of a $\mbox{P\'olya-Gamma}(b,c)$ random variable.

\subsubsection{A Decomposition of the General P\'olya Gamma Density Function}

Let $\pDensPG(\,\cdot\,;b,c)$ denote the density function of a 
$\mbox{P\'olya-Gamma}(b,c)$ random variable. 
Then for all $x,b>0$ and $c\in\real$:
\begin{equation}
\pDensPG(x;b,c)=\cosh^b(c/2)\exp\left(-\smhalf c^2x\right)\pDensPG(x;b,0).
\label{eq:PGdensResult}
\end{equation}
This result corresponds to equation (5) of Polson \textit{et al.} (2013).

\subsubsection{The Mean of a P\'olya Gamma Random Variable}

From Section 2.3 of Polson \textit{et al.} (2013),
if 
\begin{equation}
x\sim \PolyaGamma(b,c)\quad\mbox{then}\quad
E(x)=\frac{b}{2c}\tanh(c/2)=2b\lambdaJJ(c).
\label{eq:PGmean}
\end{equation}

\section{Impracticality of Ordinary Mean Field Variational Bayes}\label{sec;ordMFVB}

The impracticality of ordinary mean field variational Bayes
for model (\ref{eq:negBinModel}) stems from the following fact 
mentioned in Section \ref{sec:PGdist}: P\'olya Gamma density
functions do not admit closed forms. In this section we provide
relevant details.

From Figure \ref{fig:nbcDAG}, the Markov blanket of $\kappa$
is $\{\by,\bbeta,\bu,\balpha\}$. Therefore
$$\pDens(\kappa|\mbox{rest})=\pDens(\kappa|\by,\bbeta,\bu,\balpha)
\propto\pDens(\by|\bbeta,\bu,\kappa)
\pDens(\balpha|\by,\bbeta,\bu,\kappa)\pDens(\kappa).
$$
The factors $\pDens(\by|\bbeta,\bu,\kappa)$ and $\pDens(\kappa)$
have simple closed form expressions, but 
$$\pDens(\balpha|\by,\bbeta,\bu,\kappa)
=\prod_{i=1}^n\pDensPG\big(\alpha_i;
y_i+\kappa,(\bX\bbeta+\bZ\bu)_i-\log(\kappa)\big)
$$
depends on intractable P\'olya-Gamma density functions.
This hinders practical mean field variational Bayes for model 
(\ref{eq:negBinModel}).

\section{The Structured Mean Field Variational Bayes Alternative}

The structured mean field variational Bayes alternative goes back
to machine learning articles such as Saul \myand Jordan (1996) and 
Jaakkola (2001) with application to, for example,
coupled hidden Markov models.
Section 3.1 of Wand \textit{et al.} (2011) describes a
structured mean field variational Bayes paradigm 
for Bayesian hierarchical models within the field of statistics.
The derivations and subsequent optimal $q$-density formulae given 
in Section 3.1 of Wand \textit{et al.} (2011) are the 
basis for Algorithm \ref{alg:BCHalgo}. 

\section{Algorithm \ref{alg:BCHalgo} Justification}

We now justify the steps given in Algorithm \ref{alg:BCHalgo}.
There are two main components: (1) the optimal $\qDens$-density
derivations which lead to the Algorithm \ref{alg:BCHalgo}'s
coordinate ascent scheme and (2) the explicit expression
for the marginal log-likelihood conditional on $\kappa$,
which form the basis for the structured mean field variational
Bayes posterior density approximations.

Throughout this section we let `rest' denote all 
random variable in the model other then the random vector
of current interest. Also, $E_{\qDens(-\btheta)}$ 
signifies expectation with respect to the joint
$\qDens$-density function of all model parameters but with
$\btheta$ omitted.

\subsection{Optimal $\qDens$-Density Derivations}

We now provide the derivation of each of the optimal $\qDens$-density
functions.

\vskip2mm\noindent
\underline{\textit{Derivation of $\qDens^*(\balpha|\kappa)$}}

\vskip3mm
It follows from the second line of (\ref{eq:negBinModel}) that 
$$
\log\{\pDens(\balpha|\mbox{rest})\}
=\sumin\log\big\{\pDensPG\big(\alpha_i;y_i+\kappa, 
(\bX\bbeta+\bZ\bu)_i-\log(\kappa)\big)\big\}.
$$
In view of (\ref{eq:PGdensResult}) we then have
{\setlength\arraycolsep{1pt}
\begin{eqnarray*}
\log\{\pDens(\alpha_i|\mbox{rest})\}&=&
\log\big\{\pDensPG\big(\alpha_i;y_i+\kappa, 
(\bX\bbeta+\bZ\bu)_i-\log(\kappa)\big)\big\}\\[1ex]
&=&-\smhalf \big\{(\bX\bbeta+\bZ\bu)_i-\log(\kappa)\big\}^2\alpha_i+
\log\{\pDensPG(\alpha_i; y_i+\kappa,0)\}+\const
\end{eqnarray*}
}
where `const' denotes terms that do not depend on $\alpha_i$.
Therefore,
{\setlength\arraycolsep{1pt}
\begin{eqnarray*}
E_{\qDens(-(\alpha_i,\kappa))}\big[\log\{\pDens(\alpha_i|\mbox{rest})\}\big]
&=&-\smhalf E_{\qDens(\bbeta,\bu|\kappa)}
\big[\big\{(\bX\bbeta+\bZ\bu)_i-\log(\kappa)\big\}^2\big]\alpha_i\\[1ex]
&&\qquad+\log\{\pDensPG(\alpha_i; y_i+\kappa,0)\}+\const.
\end{eqnarray*}
}
It quickly follows that, for each $1\le i\le n$,
$$\qDens^*(\alpha_i|\kappa)\ \ \mbox{is the}\ \ 
\mbox{P\'olya-Gamma}\big(y_i+\kappa,c_{\qDens(\alpha_i|\kappa)}\big)
\ \ \mbox{density function}
$$
where
$$c_{\qDens(\alpha_i|\kappa)}\equiv
\sqrt{E_{\qDens(\bbeta,\bu|\kappa)}
\big[\big\{(\bX\bbeta+\bZ\bu)_i-\log(\kappa)\big\}^2\big]}
$$
and that 
$$\qDens^*(\balpha|\kappa)=\prod_{i=1}^n \qDens^*(\alpha_i|\kappa).$$
Next note that
{\setlength\arraycolsep{1pt}
\begin{eqnarray*}
E_{\qDens(\bbeta,\bu|\kappa)}\big[\big\{(\bX\bbeta+\bZ\bu)_i-\log(\kappa)\big\}^2
&=&\Var_{\qDens(\bbeta,\bu|\kappa)}\big(\bC [\bbeta^T\ \bu^T]^T)_i\big)\\
&&\qquad+\big\{\big(\bC E_{\qDens(\bbeta,\bu|\kappa)}[\bbeta^T\ \bu^T]^T\big)_i
-\log(\kappa)\big\}^2\\[1ex]
&=&\big(\bC\bSigma_{\qDens(\bbeta,\bu|\kappa)}\bC^T\big)_{ii}
+\big\{\big(\bC\bmu_{\qDens(\bbeta,\bu|\kappa)}\big)_i-\log(\kappa)\big\}^2\\[1ex]
&=&\Big(\diagonal\big(\bC\bSigma_{\qDens(\bbeta,\bu|\kappa)}\bC^T\big)
+\big(\bC\bmu_{\qDens(\bbeta,\bu|\kappa)}-\log(\kappa)\bone\big)^2\Big)_i.
\end{eqnarray*}
}
Hence 
$$c_{\qDens(\alpha_i|\kappa)}=
\sqrt{
\Big(\diagonal\big(\bC\bSigma_{\qDens(\bbeta,\bu|\kappa)}\bC^T\big)
+\big(\bC\bmu_{\qDens(\bbeta,\bu|\kappa)}-\log(\kappa)\bone\big)^2\Big)_i}\,.
$$
From (\ref{eq:PGmean}), 
$$\mu_{\qDens(\alpha_i)}= 2(y_i+\kappa)\lambdaJJ\big(c_{\qDens(\alpha_i|\kappa)}\big),
\quad 1\le i\le n,$$
and so 
$$\bmu_{\qDens(\balpha|\kappa)}= 2(\by+\kappa\bone)\odot
\lambdaJJ\big(c_{\qDens(\balpha|\kappa)}\big)
$$
with 
\begin{equation}
\bc_{\qDens(\balpha|\kappa)}=
\sqrt{\diagonal\big(\bC\bSigma_{\qDens(\bbeta,\bu|\kappa)}\bC^T\big)
+\big(\bC\bmu_{\qDens(\bbeta,\bu|\kappa)}-\log(\kappa)\bone\big)^2}\,.
\label{eq:cqalphaUpdate}
\end{equation}

\vskip2mm\noindent
\underline{\textit{Derivation of $\qDens^*(\bbeta,\bu|\kappa)$}}

\vskip3mm\noindent
First note that the $i$th contribution to the likelihood part of 
(\ref{eq:negBinModel}), $1\le i\le n$, can be written
{\setlength\arraycolsep{1pt}
\begin{eqnarray*}
\pDens(y_i|\bbeta,\bu,\kappa)&=&
\frac{\kappa^{\kappa}\Gamma(y_i+\kappa)\exp\big((\bX\bbeta+\bZ\bu)_i\big)^{y_i}}
{\Gamma(\kappa)\Big(\kappa+\exp\big((\bX\bbeta+\bZ\bu)_i\big)\Big)^{y_i+\kappa}\Gamma(y_i+1)}\\[1ex]
&=&\frac{\Gamma(y_i+\kappa)}{\Gamma(\kappa)\Gamma(y_i+1)}
\,\frac{\exp\big((\bX\bbeta+\bZ\bu)_i-\log(\kappa)\big)^{y_i}}
{\big\{1+\exp\big((\bX\bbeta+\bZ\bu)_i-\log(\kappa)\big)\big\}^{\kappa+y_i}}\\[1ex]
&=&\frac{\Gamma(y_i+\kappa)}{2^{y_i+\kappa}\Gamma(\kappa)\Gamma(y_i+1)}
\,\,\frac{\exp\big\{\smhalf(y_i-\kappa)\big((\bX\bbeta+\bZ\bu)_i-\log(\kappa)\big)\big\}}
{\cosh^{y_i+\kappa}\Big(\smhalf\big((\bX\bbeta+\bZ\bu)_i-\log(\kappa)\big)\Big)}.
\end{eqnarray*}
}
Then the full conditional density function of $(\bbeta,\bu)$ is
{\setlength\arraycolsep{1pt}
\begin{eqnarray*}
\pDens(\bbeta,\bu|\mbox{rest})&=&\pDens(\bbeta,\bu|\by,\balpha,\bsigsq,\kappa)
\propto\pDens(\by|\bbeta,\bu,\kappa)\pDens(\balpha|\by,\bbeta,\bu,\kappa)
\pDens(\bbeta)\pDens(\bu|\bsigsq)\\[1ex]
&=&\left(\prod_{i=1}^n\pDens(y_i|\bbeta,\bu,\kappa)
\pDensPG\big(\alpha_i;y_i+\kappa,(\bX\bbeta+\bZ\bu)_i-\log(\kappa)\big)\right) 
\pDens(\bbeta)\pDens(\bu|\bsigsq)\\[1ex]
&\propto&\Bigg(\prod_{i=1}^n  
\frac{\exp\big\{\smhalf(y_i-\kappa)\big((\bX\bbeta+\bZ\bu)_i-\log(\kappa)\big)\big\}}
{\cosh^{y_i+\kappa}\Big(\smhalf\big((\bX\bbeta+\bZ\bu)_i-\log(\kappa)\big)\Big)}\\
&&\qquad\times\cosh^{y_i+\kappa}\Big(\smhalf\big((\bX\bbeta+\bZ\bu)_i-\log(\kappa)\big)\Big)
\\
&&\qquad\times
\exp\Big[-\smhalf\big\{(\bX\bbeta+\bZ\bu)_i-\log(\kappa)\big\}^2 \alpha_i\Big]
\pDensPG(\alpha_i;y_i+\kappa,0)\Bigg)\pDens(\bbeta)\pDens(\bu|\bsigsq)
\end{eqnarray*}
}
where the last step follows from (\ref{eq:PGdensResult}). We then have
{\setlength\arraycolsep{1pt}
\begin{eqnarray*}
\pDens(\bbeta,\bu|\mbox{rest})&\propto&\left(\prod_{i=1}^n\exp\Big[\smhalf(y_i-\kappa)
(\bX\bbeta+\bZ\bu)_i
-\smhalf\big\{(\bX\bbeta+\bZ\bu)_i-\log(\kappa)\big\}^2 \alpha_i\Big]\right)\\
&&\qquad\times\exp\left(-\frac{\Vert\bbeta\Vert^2}{2\sigmabeta^2}
-\sum_{j=1}^r\frac{\Vert\bu_j\Vert^2}{2\sigma_j^2}\right)\\[1ex]
&\propto&\left(\prod_{i=1}^n\exp\Big[\smhalf(y_i-\kappa)(\bX\bbeta+\bZ\bu)_i
+\alpha_i\big\{\log(\kappa)(\bX\bbeta+\bZ\bu)_i
-\smhalf(\bX\bbeta+\bZ\bu)^2_i\big\}\Big]\right)\\
&&\qquad\times\exp\left(-\frac{\Vert\bbeta\Vert^2}{2\sigmabeta^2}
-\sum_{j=1}^r\frac{\Vert\bu_j\Vert^2}{2\sigma_j^2}\right)\\[1ex]
&=&\exp\Bigg[\big\{\smhalf(\by-\kappa\bone)+\log(\kappa)\balpha\big\}^T
\bC\bbetabuVec
            -\smhalf\bbetabuVec^T\bC^T\diag(\balpha)\bC\bbetabuVec\\
&&\qquad\qquad\qquad    -\frac{\Vert\bbeta\Vert^2}{2\sigmabeta^2}
            -\sum_{j=1}^r\frac{\Vert\bu_j\Vert^2}{2\sigma_j^2}\Bigg]\\[1ex]
&=&\exp\left\{
\left[
\begin{array}{c}
\bbetabuVec\\[3ex]
\vecof\left(\bbetabuVec\bbetabuVec^T\right)
\end{array}
\right]^T
\left[
\begin{array}{c}
\bC^T\big\{\smhalf(\by-\kappa\bone)+\log(\kappa)\balpha\big\}\\[1ex]
-\smhalf\vecof\left(\bC^T\diag(\balpha)\bC+\bMtilde\right)
\end{array}
\right]
\right\}.
\end{eqnarray*}
}
where
$$\bMtilde\equiv \mbox{blockdiag}(\sigma_{\beta}^{-2}\bI_p,\sigma_1^{-2}
\bI_{K_1},\ldots,\sigma_r^{-2}\bI_{K_r}).
$$
Therefore, 
\begin{eqnarray*}
&&E_{\qDens(-(\bbeta,\bu,\kappa))}\big[\log\{\pDens(\bbeta,\bu|\mbox{rest})\}\big]\\[1ex]
&&\quad=
\left[
\begin{array}{c}
\bbetabuVec\\[2ex]
\vecof\left(\bbetabuVec\bbetabuVec^T\right)
\end{array}
\right]^T
\left[
\begin{array}{c}
\bC^T\big\{\smhalf(\by-\kappa\bone)+\log(\kappa)\bmu_{\qDens(\balpha|\kappa)}\big\}\\[1ex]
-\smhalf\vecof\left(\bC^T\diag\big(\bmu_{\qDens(\balpha|\kappa)}\big)\bC
+\bM_{\qDens(1/\bsigsq|\kappa)}\right)
\end{array}
\right]
+\mbox{const}
\end{eqnarray*}
where $\bmu_{\qDens(\balpha|\kappa)}$ denotes the mean of the 
optimal $\qDens$-density of $\balpha|\kappa$
and `const' denotes terms that do not depend on $(\bbeta,\bu)$.
It follows quickly that
$$\qDens^*(\bbeta,\bu|\kappa)\ \mbox{is the}\ 
N\big(\bmu_{\qDens(\bbeta,\bu|\kappa)},\bSigma_{\qDens(\bbeta,\bu|\kappa)} \big)
\ \mbox{density function}
$$
where
$$\bSigma_{\qDens(\bbeta,\bu|\kappa)}
\equiv 
\left\{\bC^T\diag\big(\bmu_{\qDens(\balpha|\kappa)}\big)\bC+\bM_{\qDens(1/\bsigsq)}\right\}^{-1}
$$
and
$$
\bmu_{\qDens(\bbeta,\bu|\kappa)}\equiv
\bSigma_{\qDens(\bbeta,\bu|\kappa)}
\bC^T\big\{\smhalf(\by-\kappa\bone)+\log(\kappa)\bmu_{\qDens(\balpha|\kappa)}\big\}.
$$

\vskip2mm\noindent
\underline{\textit{Derivation of $\qDens^*(\bsigsq|\kappa)$}}

\vskip3mm
Arguments similar to those given in Appendix C of Wand \myand Ormerod (2011)
lead to 
$$\qDens^*(\bsigsq|\kappa)=\prod_{j=1}^r \qDens^*(\sigma^2_j|\kappa)$$
where, for $1\le j\le r$,
$$\qDens^*(\sigma^2_j|\kappa)
\quad\mbox{is the}\quad
\mbox{Inverse-Gamma}\Big(\smhalf(K_j+1),\mu_{\qDens(1/a_j|\kappa)}
+\smhalf\big\{\Vert\bmu_{\qDens(\bu_j|\kappa)}\Vert^2
+\tr(\bSigma_{\qDens(\bu_j|\kappa)})\big\}\Big)
$$
density function. Note that $\mu_{\qDens(1/a_j|\kappa)}$
is the mean of $1/a_j$ according to the optimal $\qDens^*$-density
described next. Also, $\bmu_{\qDens(\bu_j|\kappa)}$ and 
$\bSigma_{\qDens(\bu_j|\kappa)}$ are the sub-matrices of
$\bmu_{\qDens(\bbeta,\bu|\kappa)}$ and $\bSigma_{\qDens(\bbeta,\bu|\kappa)}$
according to the partition of $\bu$ given at 
(\ref{eq:uparticDefn}). The reciprocal moment of $\sigma^2_j$, 
according to
$\qDens^*(\sigma^2_j|\kappa)$, is
$$\mu_{\qDens(1/\sigma^2_j|\kappa)}
=\frac{K_j+1}
{2\mu_{\qDens(1/a_j|\kappa)}
+\Vert\bmu_{\qDens(\bu_j|\kappa)}\Vert^2
+\tr(\bSigma_{\qDens(\bu_j|\kappa)})}.
$$

\vskip2mm\noindent
\underline{\textit{Derivation of $\qDens^*(\ba|\kappa)$}}

\vskip3mm
Steps provided by Appendix C of Wand \myand Ormerod (2011)
lead to 
$$\qDens^*(\ba|\kappa)=\prod_{j=1}^r \qDens^*(a_j|\kappa)$$
where, for $1\le j\le r$,
$$\qDens^*(a_j|\kappa)
\quad\mbox{is the}\quad
\mbox{Inverse-Gamma}\Big(1,\mu_{\qDens(1/\sigma^2_j|\kappa)}
+\ssigma^{-2}\Big)
$$
density function. The reciprocal moment of $a_j$, according to
$\qDens^*(a_j|\kappa)$, is
$$\mu_{\qDens(1/a_j|\kappa)}
=\frac{1}{\mu_{\qDens(1/\sigma^2_j|\kappa)}+\ssigma^{-2}}.
$$

\subsection{The Approximate Marginal Log-Likelihood}

The approximate marginal log-likelihood, conditional on $\kappa$, is
{\setlength\arraycolsep{1pt}
\begin{eqnarray*}
\log\{\pDensUnder(\by|\kappa)\}
&=&E_{\qDens(-\kappa)}\big[\log\{\pDens(\by,\balpha,\bbeta,\bu,\bsigsq,\ba|\kappa)\}
-\log\{\qDens(\balpha,\bbeta,\bu,\bsigsq,\ba|\kappa)\}\big]\\[1ex]
&=&E_{\qDens(-\kappa)}\big[\log\{\pDens(\by|\bbeta,\bu,\kappa)\pDens(\balpha|\by,\bbeta,\bu,\kappa)
\pDens(\bbeta,\bu|\bsigsq,\kappa)\pDens(\bsigsq|\ba,\kappa)\pDens(\ba|\kappa)\}\\[1ex]
&&\qquad-\log\{\qDens(\balpha|\kappa)\qDens(\bbeta,\bu|\kappa)\qDens(\bsigsq|\kappa)
\qDens(\ba|\kappa)\}\big]\\[1ex]
&=&E_{\qDens(-\kappa)}\big[\log\{\pDens(\by|\bbeta,\bu,\kappa)\}
+\log\{\pDens(\balpha|\by,\bbeta,\bu,\kappa)\}-\log\{\qDens(\balpha|\kappa)\}\\[1ex]
&&\quad
+\log\{\pDens(\bbeta,\bu|\bsigsq,\kappa)\}-\log\{\qDens(\bbeta,\bu|\kappa)\}
+\log\{\pDens(\bsigsq|\ba,\kappa)\}-\log\{\qDens(\bsigsq|\kappa)\}\\[1ex]
&&\quad+\log\{\pDens(\ba|\kappa)\}-\log\{\qDens(\ba|\kappa)\}\big].
\end{eqnarray*}
}

\noindent
\underline{\textit{Simplification of $E_{\qDens(-\kappa)}\big[\log\{\pDens(\by|\bbeta,\bu,\kappa)\}\big]$}}

\noindent\vskip3mm
Since
$$
\pDens(y_i|\bbeta,\bu,\kappa)=
\frac{\Gamma(y_i+\kappa)}{2^{y_i+\kappa}\Gamma(\kappa)\Gamma(y_i+1)}
\,\,\frac{\exp\big\{\smhalf(y_i-\kappa)\big((\bX\bbeta+\bZ\bu)_i-\log(\kappa)\big)\big\}}
{\cosh^{y_i+\kappa}\Big(\smhalf\big((\bX\bbeta+\bZ\bu)_i-\log(\kappa)\big)\Big)}
$$
we have
{\setlength\arraycolsep{1pt}
\begin{eqnarray*}
\log\{\pDens(\by|\bbeta,\bu,\kappa)\}
&=&\sumin\Big[\log\{\Gamma(y_i+\kappa)\}
-(y_i+\kappa)\log(2)-\log\{\Gamma(y_i+1)\}\\[1ex]
&&\qquad+\smhalf(y_i-\kappa)\big((\bX\bbeta+\bZ\bu)_i-\log(\kappa)\big)\\[1ex]
&&\qquad-(y_i+\kappa)\log\left\{
\cosh\Big(\smhalf\big((\bX\bbeta+\bZ\bu)_i-\log(\kappa)\big)\Big)\right\}\Big]
-n \log\{\Gamma(\kappa)\}.
\end{eqnarray*}
}
This leads to 
{\setlength\arraycolsep{1pt}
\begin{eqnarray*}
E_{\qDens(-\kappa)}\big[\log\{\pDens(\by|\bbeta,\bu,\kappa)\}\big]
&=&-(\by^T\bone)\log(2)-\bone^T\log\{\Gamma(\by+\bone)\}
+\bone^T\log\{\Gamma(\by+\kappa\bone)\}\\[1ex]
&&\quad+n\big[\smhalf\kappa\log(\kappa)-\log(2)\kappa-\log\{\Gamma(\kappa)\}\big]\\[1ex]
&&\quad -\smhalf(\by^T\bone)\log(\kappa)
+\smhalf\bmu_{\qDens(\bbeta,\bu|\kappa)}^T\big(\bC^T\by-\kappa\,\bC^T\bone\big)\\[1ex]
&&\quad-\sumin E_{\qDens(-\kappa)}\left[(y_i+\kappa)\log\left\{
\cosh\Big(\smhalf\big((\bX\bbeta+\bZ\bu)_i-\log(\kappa)\big)\Big)\right\}\right].
\end{eqnarray*}
}

\noindent
\underline{\textit{Simplification of 
$E_{\qDens(-\kappa)}\big[\log\{\pDens(\balpha|\by,\bbeta,\bu,\kappa)\}-\log\{\qDens(\balpha|\kappa)\}\big]$}}

\vskip3mm
First note that 
\begin{eqnarray*}
&&\log\{\pDens(\balpha|\by,\bbeta,\bu,\kappa)\}-\log\{\qDens(\balpha|\kappa)\}\\[1ex]
&&\qquad
=\sumin\Big[\log\big\{\pDensPG(\alpha_i;y_i+\kappa,(\bX\bbeta+\bZ\bu)_i-\log(\kappa))\big\}
-\log\big\{\pDensPG(\alpha_i;y_i+\kappa,c_{\qDens(\alpha_i|\kappa)})
\big\}\Big].
\end{eqnarray*}
Since 
\begin{eqnarray*}
&&
\log\{\pDensPG(\alpha_i;y_i+\kappa,(\bX\bbeta+\bZ\bu)_i-\log(\kappa))\}\\
&&\qquad=(y_i+\kappa)\log\left\{
\cosh\Big(\smhalf\big((\bX\bbeta+\bZ\bu)_i-\log(\kappa)\big)\Big)\right\}
-\smhalf\big((\bX\bbeta+\bZ\bu)_i-\log(\kappa)\big)^2\alpha_i\\
&&\qquad\qquad\qquad-\log\{\pDensPG(\alpha_i;y_i+\kappa,0)\}
\end{eqnarray*}
and
\begin{eqnarray*}
&&
\log\{\pDensPG(\alpha_i;y_i+\kappa,c_{\qDens(\alpha_i|\kappa)}\}\\
&&\qquad=(y_i+\kappa)\log\left\{
\cosh\Big(\smhalf c_{\qDens(\alpha_i|\kappa)} \Big)\right\}
-\smhalf c_{\qDens(\alpha_i|\kappa)}^2\alpha_i-\log\{\pDensPG(\alpha_i;y_i+\kappa,0)\}
\end{eqnarray*}
we get the cancellation of the
$\log\{\pDensPG(\alpha_i;y_i+\kappa,0)\}$ terms and the following 
explicit expression for the log-density difference:
\begin{eqnarray*}
&&
\log\big\{\pDensPG(\alpha_i;y_i+\kappa,(\bX\bbeta+\bZ\bu)_i-\log(\kappa))\big\}
-\log\big\{\pDensPG(\alpha_i;y_i+\kappa,c_{\qDens(\alpha_i|\kappa)})
\big\}\\[1ex]
&&\quad=(y_i+\kappa)\log\left\{
\cosh\Big(\smhalf\big((\bX\bbeta+\bZ\bu)_i-\log(\kappa)\big)\Big)\right\}
-(y_i+\kappa)\log\left\{
\cosh\Big(\smhalf c_{\qDens(\alpha_i|\kappa)} \Big)\right\}\\
&&\qquad\qquad+\smhalf\alpha_i\Big\{ c_{\qDens(\alpha_i|\kappa)}^2\
-\Big((\bX\bbeta+\bZ\bu)_i-\log(\kappa)\Big)^2\Big\}.
\end{eqnarray*}
The cancellation of the $\log\{\pDensPG(\alpha_i;y_i+\kappa,0)\}$ terms 
is very important from a practical standpoint since 
$\pDensPG(\alpha_i;y_i+\kappa,0)$ does not admit a closed form.

We then have
\begin{eqnarray*}
&&E_{\qDens(-\kappa)}\big[\log\{\pDens(\balpha|\by,\bbeta,\bu,\kappa)\}-\log\{\qDens(\balpha|\kappa)\}
\big]
\\[1ex]
&&\qquad=\sumin E_{\qDens(-\kappa)}\Big[(y_i+\kappa)\log\left\{
\cosh\Big(\smhalf\big((\bX\bbeta+\bZ\bu)_i-\log(\kappa)\big)\Big)\right\}\Big]
\\
&&\qquad\qquad
-(\by+\kappa\bone)^T\log\left\{
\cosh\Big(\smhalf c_{\qDens(\balpha|\kappa)}\Big)\right\}
+\smhalf\bc_{\qDens(\balpha|\kappa)}^T\diag\{\mu_{\qDens(\balpha|\kappa)}\}
\bc_{\qDens(\balpha|\kappa)}\\[1ex]
&&\qquad\qquad-\smhalf\sumin E_{\qDens(-\kappa)}
\Big\{\alpha_i\big((\bX\bbeta+\bZ\bu)_i-\log(\kappa)\big)^2\Big\}.
\end{eqnarray*}

\noindent
\underline{\textit{Simplification of 
$E_{\qDens(-\kappa)}\big[\log\{\pDens(\bbeta,\bu|\bsigsq,\kappa)\}-\log\{\qDens(\bbeta,\bu|\kappa)\}\big]$}}

\vskip3mm
For this contribution, we have
{\setlength\arraycolsep{1pt}
\begin{eqnarray*}
\log\{\pDens(\bbeta,\bu|\bsigsq,\kappa)\}-\log\{\qDens(\bbeta,\bu|\kappa)\}
&=&-\smhalf p\log(\sigsqbeta) -\smhalf\sum_{j=1}^rK_j\log(\sigma_j^2)\\[2ex]
&&\quad-\frac{\Vert\bbeta\Vert^2}{2\sigsqbeta}
-\sum_{j=1}^r\frac{\Vert\bu_j\Vert^2}{2\sigma^2_j}
+\smhalf\log|\bSigma_{\qDens(\bbeta,\bu|\kappa)}|\\[1ex]
&&\quad+\smhalf\left(\left[\begin{array}{c}
\bbeta\\
\bu
\end{array}
\right]-\bmu_{\qDens(\bbeta,\bu|\kappa)}\right)^T 
\bSigma_{\qDens(\bbeta,\bu|\kappa)}^{-1}
\left(\left[\begin{array}{c}
\bbeta\\
\bu
\end{array}
\right]-\bmu_{\qDens(\bbeta,\bu|\kappa)}\right).
\end{eqnarray*}
}
We then obtain
{\setlength\arraycolsep{1pt}
\begin{eqnarray*}
&&E_{\qDens(-\kappa)}\big[\log\{\pDens(\bbeta,\bu|\bsigsq,\kappa)\}-\log\{\qDens(\bbeta,\bu|\kappa)\}\big]\\[1ex]
&&\qquad\qquad\qquad=-\smhalf p\log(\sigsqbeta) -\smhalf\sum_{j=1}^rK_j
E_{\qDens(-\kappa)}\{\log(\sigma_j^2)\}\\[2ex]
&&\quad\qquad\qquad\qquad
-\frac{\Vert\bmu_{\qDens(\bbeta,\bu|\kappa)}\Vert^2+\tr(\bSigma_{\qDens(\bbeta,\bu|\kappa)})}{2\sigsqbeta}
-\smhalf\sum_{j=1}^r\mu_{\qDens(1/\sigma_j^2|\kappa)}
\big\{\Vert\bmu_{\qDens(\bu_j|\kappa)}\Vert^2
+\tr\big(\bSigma_{\qDens(\bu_j|\kappa)}\big)\big\}
\\[1ex]
&&\quad\qquad\qquad\qquad +\smhalf\log|\bSigma_{\qDens(\bbeta,\bu|\kappa)}|
+\smhalf p+\smhalf\sum_{j=1}^r K_j.
\end{eqnarray*}
}

\noindent
\underline{\textit{Simplification of 
$\log\{\pDens(\bsigsq|\ba,\kappa)\}-\log\{\qDens(\bsigsq|\kappa)\}$}}

\vskip3mm
\noindent
Simple manipulations give
{\setlength\arraycolsep{1pt}
\begin{eqnarray*}
\log\{\pDens(\bsigsq|\ba,\kappa)\}-\log\{\qDens(\bsigsq|\kappa)\}
&=&\sum_{j=1}^r\Big[-\smhalf\log(a_j)-\smhalf\log(\pi)
-1/(\sigma^2_ja_j)\\[2ex]
&&\quad-\smhalf(K_j+1)
\log\big(\lambda_{\qDens(\sigma^2_j|\kappa)}\big)
+\log\big\{\Gamma\big(\smhalf(K_j+1)\big)\big\}\\[1ex]
&&\quad +\smhalf\,K_j\log(\sigma^2_j)
+\lambda_{\qDens(\sigma^2_j|\kappa)}\big/\sigma^2_j\Big]
\end{eqnarray*}
}
where 
$$\lambda_{\qDens(\sigma^2_j|\kappa)}\equiv 
\mu_{\qDens(1/a_j|\kappa)}
+\smhalf\big\{\Vert\bmu_{\qDens(\bu_j|\kappa)}\Vert^2
+\tr(\bSigma_{\qDens(\bu_j|\kappa)})\big\}.
$$
Therefore
{\setlength\arraycolsep{1pt}
\begin{eqnarray*}
E_{\qDens(-\kappa)}\big[\log\{\pDens(\bsigsq|\ba,\kappa)\}-\log\{\qDens(\bsigsq|\kappa)\}\big]
&=&-\smhalf r\log(\pi) +\sum_{j=1}^r
\log\big\{\Gamma\big(\smhalf(K_j+1)\big)\big\}\\[1ex]
&&\quad-\smhalf\sum_{j=1}^r E_{\qDens(-\kappa)}\{\log(a_j)\}
+\smhalf\sum_{j=1}^rK_jE_{\qDens(-\kappa)}\{\log(\sigma^2_j)\}
\\[1ex]
&&\quad
+\sum_{j=1}^r\Big[
\lambda_{\qDens(\sigma^2_j|\kappa)}\mu_{\qDens(1/\sigma^2_j|\kappa)}
-\mu_{\qDens(1/\sigma^2_j|\kappa)}\mu_{\qDens(1/a_j|\kappa)}\\
&&\quad\qquad
-\smhalf(K_j+1)\log\big(\lambda_{\qDens(\sigma^2_j|\kappa)}\big)\Big].
\end{eqnarray*}
}

\vskip2mm
\noindent
\underline{\textit{Simplification of 
$E_{\qDens(-\kappa)}\big[\log\{\pDens(\ba|\kappa)\}-\log\{\qDens(\ba|\kappa)\}\big]$}}

\vskip3mm\noindent
Lastly, we have
{\setlength\arraycolsep{1pt}
\begin{eqnarray*}
\log\{\pDens(\ba|\kappa)\}-\log\{\qDens(\ba|\kappa)\}
&=&\sum_{j=1}^r\Big[-\log(\ssigma)-\smhalf\log(\pi)
+\smhalf\log(a_j)
-1/(a_j\ssigma^2)\\[1ex]
&&\qquad-\log\big(\lambda_{\qDens(a_j|\kappa)}\big)
+\lambda_{\qDens(a_j|\kappa)}\mu_{\qDens(1/a_j|\kappa)}\Big]
\end{eqnarray*}
}
where
$$\lambda_{\qDens(a_j|\kappa)}\equiv 
\mu_{\qDens(1/\sigma^2_j|\kappa)}+\ssigma^{-2}.$$
The required $\qDens$-density expectation is
{\setlength\arraycolsep{1pt}
\begin{eqnarray*}
E_{\qDens(-\kappa)}\big[\log\{\pDens(\ba|\kappa)\}-\log\{\qDens(\ba|\kappa)\}\big]
&=&-r\log(\ssigma)-\smhalf\,r\log(\pi)
+\smhalf\sum_{j=1}^r E_{\qDens(-\kappa)}\{\log(a_j)\}\\[0ex]
&&\quad+\sum_{j=1}^r\Big[\lambda_{\qDens(a_j|\kappa)}
\mu_{\qDens(1/a_j|\kappa)}-\mu_{\qDens(1/a_j|\kappa)}\big/\ssigma^2
-\log\big(\lambda_{\qDens(a_j|\kappa)}\big)\Big].
\end{eqnarray*}
}

\vskip2mm
\noindent
\underline{\textit{Fully Simplified $\log\{\pDensUnder(\by|\kappa)\}$ Expression}}

\vskip3mm
Combining each of the simplified contributions, we obtain
{\setlength\arraycolsep{1pt}
\begin{eqnarray*}
\log\{\pDensUnder(\by|\kappa)\}&=&
\bone^T\log\{\Gamma(\by+\kappa\bone)\}
+n\big[\smhalf\kappa\log(\kappa)-\log(2)\kappa-\log\{\Gamma(\kappa)\}\big]\\[1ex]
&&\qquad -\smhalf(\by^T\bone)\log(\kappa)
+\smhalf\bmu_{\qDens(\bbeta,\bu|\kappa)}^T\big(\bC^T\by-\kappa\,\bC^T\bone\big)\\[1ex]
&&\qquad -(\by+\kappa\bone)^T\log\left\{
\cosh\Big(\smhalf c_{\qDens(\balpha|\kappa)}\Big)\right\}
-\frac{\Vert\bmu_{\qDens(\bbeta|\kappa)}\Vert^2+\tr(\bSigma_{\qDens(\bbeta|\kappa)})}{2\sigsqbeta}
\\
&&\qquad
-\smhalf\sum_{j=1}^r\mu_{\qDens(1/\sigma_j^2|\kappa)}
\big\{\Vert\bmu_{\qDens(\bu_j|\kappa)}\Vert^2
+\tr\big(\bSigma_{\qDens(\bu_j|\kappa)}\big)\big\}
+\smhalf\log|\bSigma_{\qDens(\bbeta,\bu|\kappa)}|\\[1ex]
&&\qquad
+\sum_{j=1}^r\Big\{
\lambda_{\qDens(\sigma^2_j|\kappa)}\mu_{\qDens(1/\sigma^2_j|\kappa)}
-\mu_{\qDens(1/\sigma^2_j|\kappa)}\mu_{\qDens(1/a_j|\kappa)}
-\smhalf(K_j+1)\log\big(\lambda_{\qDens(\sigma^2_j|\kappa)}\big)\\[1ex]
&&\qquad\qquad\quad+\lambda_{\qDens(a_j|\kappa)}
\mu_{\qDens(1/a_j|\kappa)}-\mu_{\qDens(1/a_j|\kappa)}\big/\ssigma^2
-\log\big(\lambda_{\qDens(a_j|\kappa)}\big)\Big\}\\
\end{eqnarray*}
}

%
%

%
{\setlength\arraycolsep{1pt}
\begin{eqnarray*}
\qquad\qquad\quad\quad&&\qquad
+\smhalf\bc_{\qDens(\balpha|\kappa)}^T\diag\{\mu_{\qDens(\balpha|\kappa)}\}
\bc_{\qDens(\balpha|\kappa)}
-\smhalf\sumin E_{\qDens(-\kappa)}\Big\{\big((\bX\bbeta+\bZ\bu)_i-\log(\kappa)\big)^2\Big\}\\[1ex]
&&\qquad -(\by^T\bone)\log(2)-\bone^T\log\{\Gamma(\by+\bone)\}
-\smhalf p\log(\sigsqbeta)+\smhalf p+\smhalf\sum_{j=1}^r K_j\\[0ex]
&&\qquad -r\log(\pi) -r\log(\ssigma)+\sum_{j=1}^r
\log\big\{\Gamma\big(\smhalf(K_j+1)\big)\big\}
\end{eqnarray*}
}
which leads to the expression
{\setlength\arraycolsep{1pt}
\begin{eqnarray*}
\log\{\pDensUnder(\by|\kappa)\}&=&\logML(\kappa)
+\smhalf \bc_{\qDens(\balpha|\kappa)}^T\diag\{\mu_{\qDens(\balpha|\kappa)}\}
\bc_{\qDens(\balpha|\kappa)}
-\smhalf\sumin E_{\qDens(-\kappa)}
\Big\{\alpha_i\big((\bX\bbeta+\bZ\bu)_i-\log(\kappa)\big)^2\Big\}\\[1ex]
&&\qquad+\const
\end{eqnarray*}
}
where `const' denotes terms that do not involve $\kappa$ or $\qDens$-density 
parameters and
{\setlength\arraycolsep{1pt}
\begin{eqnarray*}
\logML(\kappa)&\equiv&
\bone^T\log\{\Gamma(\by+\kappa\bone)\}
+n\big[\smhalf\kappa\log(\kappa)-\log(2)\kappa-\log\{\Gamma(\kappa)\}\big]\\[1ex]
&&\qquad -\smhalf(\by^T\bone)\log(\kappa)
+\smhalf\bmu_{\qDens(\bbeta,\bu|\kappa)}^T\big(\bC^T\by-\kappa\,\bC^T\bone\big)\\[1ex]
&&\qquad -(\by+\kappa\bone)^T\log\left\{
\cosh\Big(\smhalf c_{\qDens(\balpha|\kappa)}\Big)\right\}
-\frac{\Vert\bmu_{\qDens(\bbeta|\kappa)}\Vert^2+\tr(\bSigma_{\qDens(\bbeta|\kappa)})}{2\sigsqbeta}
\\
&&\qquad
-\smhalf\sum_{j=1}^r\mu_{\qDens(1/\sigma_j^2|\kappa)}
\big\{\Vert\bmu_{\qDens(\bu_j|\kappa)}\Vert^2
+\tr\big(\bSigma_{\qDens(\bu_j|\kappa)}\big)\big\}
+\smhalf\log|\bSigma_{\qDens(\bbeta,\bu|\kappa)}|\\[1ex]
&&\qquad
+\sum_{j=1}^r\Big\{
\lambda_{\qDens(\sigma^2_j|\kappa)}\mu_{\qDens(1/\sigma^2_j|\kappa)}
-\mu_{\qDens(1/\sigma^2_j|\kappa)}\mu_{\qDens(1/a_j|\kappa)}
-\smhalf(K_j+1)\log\big(\lambda_{\qDens(\sigma^2_j|\kappa)}\big)\\[1ex]
&&\qquad\qquad\quad+\lambda_{\qDens(a_j|\kappa)}
\mu_{\qDens(1/a_j|\kappa)}-\mu_{\qDens(1/a_j|\kappa)}\big/\ssigma^2
-\log\big(\lambda_{\qDens(a_j|\kappa)}\big)\Big\}.
\end{eqnarray*}
}
Lastly, we note that the  
$$\bc_{\qDens(\balpha|\kappa)}^T\diag\{\mu_{\qDens(\balpha|\kappa)}\}
\bc_{\qDens(\balpha|\kappa)}
\quad\mbox{and}\quad
\sumin E_{\qDens(-\kappa)}\Big\{\alpha_i\big((\bX\bbeta+\bZ\bu)_i-\log(\kappa)\big)^2\Big\}
$$
terms cancel with each other in the $\qDens$-density updates, and we have the 
simpler marginal log-likelihood expression
$$\log\{\pDensUnder(\by|\kappa)\}=\logML(\kappa)+\const.$$

\section*{Additional References}

\bib
Jaakkola, T.S. (2001). Tutorial on variational approximation
methods. In Opper, M. \myand Saad, D., editors,
\textit{Advanced Mean Field Methods: Theory and Practice},
129--160. Cambridge, Massachusetts: MIT Press.

\bib
Wand, M.P. and Ormerod, J.T. (2011).
Penalized wavelets: embedding wavelets
into semiparametric regression.
{\it Electronic Journal of Statistics},
{\bf 5}, 1654--1717.

\end{document}